\documentclass[pra,aps,amsmath,amssymb,amsfonts,
twocolumn,nofootinbib]{revtex4}
\usepackage{bm,mathrsfs}
\usepackage{graphicx}
\usepackage{bbm}
\usepackage{times}
\usepackage{verbatim}
\usepackage{float}
\usepackage{placeins}
\usepackage{textgreek}
\usepackage{color}
\allowdisplaybreaks[4]
\newcommand{\imag}{i}
\newcommand{\e}{\mathrm{e}}
\usepackage{}
\usepackage{amssymb}
\usepackage{bm,mathrsfs}
\usepackage{graphicx}
\usepackage{epsfig}
\usepackage{amsmath,bbm}
\usepackage{amsfonts,amssymb}
\usepackage{times}
\usepackage{verbatim}
\usepackage[sort&compress]{natbib}
\usepackage{amsmath}
\usepackage{bm}
\usepackage{textgreek}
\usepackage{textcomp}
\allowdisplaybreaks[4]
\usepackage[colorlinks,breaklinks,linkcolor=blue,anchorcolor=blue,citecolor=blue,urlcolor=magenta]{hyperref}
\usepackage{color}
\definecolor{zzz}{rgb}{0.9,0.0,0.4}
\definecolor{zzz}{rgb}{0.9,0.0,0.4}

\begin{document}
\raggedbottom
\setlength{\parindent}{1em}
\setlength{\parskip}{0pt}

\title{Perfect Bound States and non-Markovian Dynamics for Imperfect Giant Atoms in a Semi-Infinite Waveguide}
\author{T. Y. Liu,$^{1}$ X. Y. Li,$^{1}$  W. Y. Hu,$^{1}$ T. Z. Luan,$^{2}$ Cheng Shang,$^{3,}$\footnote{\textcolor{zzz}{Corresponding author: cheng.shang@riken.jp}} and H. Z. Shen$^{1,}$\footnote{\textcolor{zzz}{Corresponding author: shenhz458@nenu.edu.cn }}}
\affiliation{$^1$Center for Quantum Sciences and School of Physics, Northeast Normal University, Changchun 130024, China\\
$^2$School of Physics Science and Technology, Shenyang Normal University, Shenyang 110034, China\\
$^3$Analytical Quantum Complexity RIKEN Hakubi Research Team,
RIKEN Center for Quantum Computing (RQC), 2-1 Hirosawa, Wako, Saitama 351-0198, Japan}

\date{\today}

\begin{abstract}
Bound states in structured reservoirs require exact interference conditions and are therefore sensitive to imperfections in the coupling geometry. In this paper, we study a two-level giant atom coupled at multiple points to a mirror-terminated semi-infinite waveguide. The exact atom-photon bound state requires both cancellation of the outgoing radiation amplitude and dispersive matching to the atomic detuning. Generalized delay-equation dynamics, continued-pole spectra, and real-space photon fields verify this criterion for coupling-strength, position, and coupling-phase imperfections. For representative realizations, retuning restores one-, two-, and three-root responses after strength errors and one- and two-root responses after position errors. The specified position-imperfect three-root configuration fails because its $2\pi$ root is no longer dark, consistent with a $58.77\%$ spectral mismatch. Ensembles of 1000 realizations at each error amplitude show near-unity target-state recovery for strength errors, unity recovery of the specified one- and two-root targets but no three-root recovery for position errors, and no nontrivial exact recovery for independent phase errors. Optimized phase-imperfect responses instead retain finite linewidths. Root matching with realization-dependent control retuning therefore constitutes a compensation procedure, rather than generic passive robustness against arbitrary disorder.
\end{abstract}

\maketitle
\section{Introduction}
Waveguide quantum electrodynamics provides a controlled setting for studying how propagation, interference, and reservoir memory reshape light-matter interactions \cite{Roy0210012017,Sheremet210306824,Breuer2016RMP,Shen2130012005}. Giant atoms extend this setting beyond the local-coupling limit because the same emitter interacts with a waveguide at several spatially separated points. Acoustic and superconducting implementations have established this nonlocal coupling architecture \cite{Gustafsson2072014,Manenti9752017,Andersson11232019,Kannan7752020}. Early giant-acoustic-atom theory identified deterministic propagation delay and frequency-dependent interference as intrinsic consequences of this geometry \cite{Guo053821017,Kockum0138372014}. Subsequent theory and experiment have shown that self-interference can tailor relaxation, Lamb shifts, level structure, decoherence-free interactions, vacuum-Rabi splitting, internal-resonator interference, Fano features, optical response, and non-dipolar photon scattering \cite{Kockum1404042018,Carollo0431842020,Vadiraj200314167,Ask0138402019,Yang1151042021,Kuo230707949,Shen2019OE,Cai0337102021,YanLu2024DarkLines,JASun2024PRA,Xin2024PRA,Yue2026PhotonInteractions,Wang2024MagnonBlockade}.

Spatially separated coupling points can also support atom-photon bound states embedded in an otherwise open waveguide. Such states appear as nondecaying or oscillating pole contributions and have been studied in giant-atom, chiral, photonic-lattice, and structured-waveguide settings \cite{Guo0430142020,Guo0337062020,Shi2016PRX,Zhao0538552020,Wang0436022021,Sundaresan2019PRX,Lim0237162023,Xiao802022,Vega0535222021,Lu2026PRA113,Li2024TunableChiralBound,Gan2025MagnonicBarnett,Schrinski2024PhotonBound,Yan2026FlatbandBIC,Zhao2024TopologicalGiant,Lin2023PRA,TLi2023PRA}. The influence of decoherence on the Hall conductance of two-band systems has also been investigated \cite{Shen2017PRE}. More generally, Hofstadter-ladder, metamaterial, topological, broken-time-reversal, and higher-dimensional reservoirs broaden the available interference landscapes for giant-atom quantum optics \cite{Wang0437032022,Gao0137162024,Wang0537202024,Cui2020PRA102,Wang0132792024,Du0450102023,Shen2018PRA98,Tudela2036032019,Zhu2025PRA111,Vega2023TopologicalMultimode,BagRoy2023StructuredWaveguides,Garmon2026WeakCouplingBIC,Wang2025ChiralTopologicalBIC,Xia2025PRA}. Resonator-array and strong-coupling studies additionally connect the long-time dynamics to atom-photon dressed states and to the number of bound eigenstates in the composite spectrum \cite{Yu2025PRA111,Jia94952024,Yu2026PRA114,WuAn0537242026,Sun0637132026,Cui2026OL,Calajo2016PRA,Shi2024MultimerStates,GSun2024FP,CLYang2023PRA}. Their formation is controlled by destructive radiation interference rather than by a finite photonic band gap alone. The same principle underlies complex decoherence-free interactions, chiral networks, nonreciprocal excitation transfer and scattering, and long-lived collective states \cite{Du0237052023,Shen2019PRA99,Soro0137102023,Soro023712,Wang0350072022,Santos0536012023,Liu2026PRA113,Weng2025PRA111,YinLiao2023Entanglement,Li2024ChiralScattering,PFWang2023PRA,PFWang2023OE}. Collisional descriptions, time-dependent couplings, transition-frequency modulation, and auxiliary resonators provide complementary routes for controlling this interference \cite{Cilluffo0430702020,Du0231982022,Du123012023,Shen2018OL,GSun2023OE,WALi2023PRA}.

The semi-infinite waveguide adds a distinct coherent-feedback channel, where atom-mirror interference and single-end waveguide studies show that radiation emitted toward the termination can return after a finite propagation time, so a Markovian decay rate cannot generally describe the full dynamics \cite{Eschner4952001,Dorner2002PRA,Tufarelli0138202013,Tufarelli0121132014,Guimond0440122017,Bradford0638302013,Guimond0238082016,ArranzRegidor2021RetardedWQED}. Related work has treated position-dependent emission and memory in semi-infinite rectangular guides, laser-tunable multilevel dynamics, non-Markovian collective emission, scattering-induced memory, and multiphoton feedback \cite{Song592018,Zeng0303052023,Xin0537062022,Sinha0436032020,Fang0430352018,Barkemeyer0237082022,Crowder0137142022,Barkemeyer0337042021,Barkemeyer19000782020,Gao2024PRA,Gu2024PRA,Gu2023PRA,Qiu2024PRR,YZZhang2023PRA,Cai2023PRA}. Delayed feedback can create persistent beats, long-distance correlations, and bound states in the continuum \cite{Milonni10961974,Zheng1136012013,Ballestero0730152013,Fang0538452015,Calajo0736012019,Dinc2132019,He2025AQT8}. In the clean semi-infinite giant-atom system, the mirror-assisted feedback produces static and oscillating bound components whose frequencies are fixed by imaginary-axis poles \cite{LiShen2024}. Related studies of nonperturbative giant-atom dynamics, topology, collective radiance, chiral networks, and limits imposed by structured photonic baths further emphasize that retardation and interference are required to be treated at the level of the full pole equation \cite{Longhi30172020,Yin0637032022,Noachtar0137022022,Brianso0637172022,Qiu2242122023,Ramos0621042016,Cui2018PRA97,Burillo0137092020,XuGuo2024CatchRelease}.

Imperfections make the bound-state question more restrictive in realistic settings. Random coupling-strength and position variations can alter decoherence-free behavior and shift or broaden bound-state-in-the-continuum features \cite{HanGuo2026,WangZhang2026PRA,Monteiro2025PRA,Li2025PRA,Lin2024PRA}. Complex local couplings add relative phases that rotate the radiation amplitudes in the complex plane and change the associated non-Markovian dynamics \cite{Sun2025,Zhu0437102022,Cui2026PRA114,Du2236022022,Zhang2025CTP77}. Existing results therefore identify several relevant control channels, but a direct criterion separating cancellation of outgoing radiation from reactive frequency matching is still needed for arbitrary positions and complex local couplings. Without that separation, a long-lived numerical plateau can be mistaken for an exact bound state, and realization-dependent parameter compensation can be mistaken for passive statistical robustness.

In this work, we derive a unified radiation-dispersion decomposition for a two-level giant atom coupled at arbitrary ordered positions to a mirror-terminated waveguide. Within this framework, we consider bound-state configurations by identifying one, two, or three distinct bound-state roots. The real part of the imaginary-axis pole is identified as the nonnegative radiation functional, whose vanishing closes the outgoing channel. The remaining imaginary part defines an independent dispersive-matching function. For configurations with multiple bound-state roots, all selected roots must satisfy the same matching conditions for common system parameters. We use this equation pair to distinguish exact matched poles from finite-linewidth transients under coupling-strength, position, and phase imperfections. The evidence combines generalized delay-equation dynamics, pole continuation, frequency-domain diagnostics, and real-space fields generated from the same pole set. 

The paper is organized as follows. In Sec.~\ref{sec:Generalized model},
we establish the generalized model and the common pole criterion.
In Secs.~\ref{sec:strength-disorder}--\ref{sec:phase-disorder},
we investigate coupling-strength, position, and coupling-phase
imperfections. In Sec.~\ref{sec:ensemble-statistics}, we compare the
three imperfection channels over disorder ensembles and identify their
failure modes. Sec.~\ref{sec:conclusion} is devoted to conclusions.
Technical derivations and numerical details are given in the Appendices.
\section{Generalized model, dynamical solution, and bound states}
\label{sec:Generalized model}
\begin{figure}[t]
\centering
\hspace*{-0.7cm}\includegraphics[width=1\columnwidth]{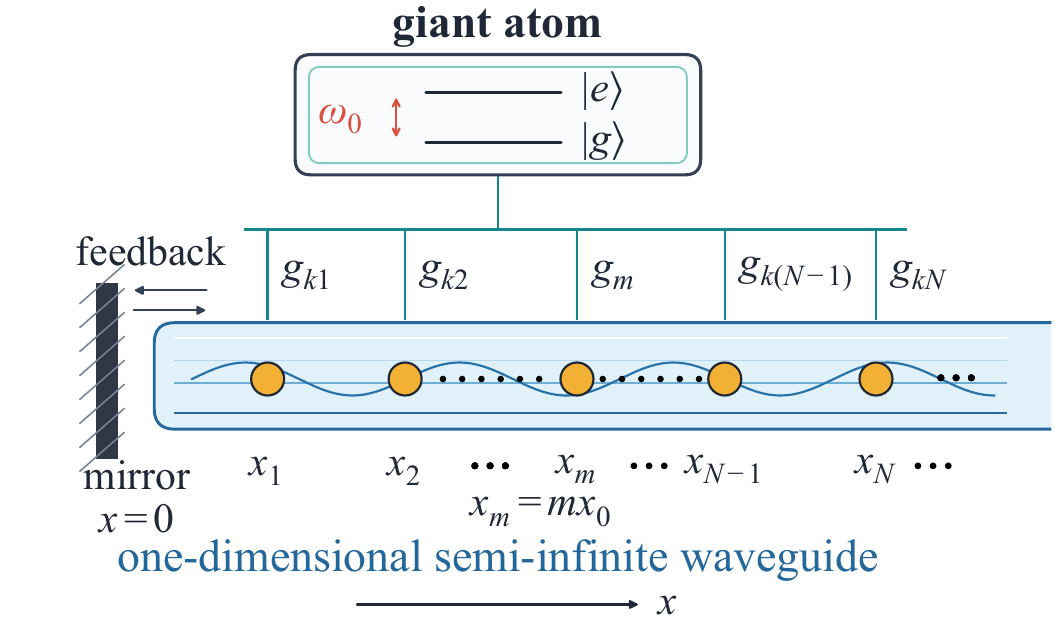}
\caption{Schematic of a two-level giant atom coupled at $N$ spatially separated points to a semi-infinite waveguide terminated by a perfect mirror at $x=0$. The atom has ground state $|g\rangle$, excited state $|e\rangle$, and transition frequency $\omega_0$, while $\mathrm{g}_{km}$ denotes the coupling to waveguide mode $k$ at the $m$th coupling point. Photons emitted into the waveguide can be reflected by the mirror and subsequently interact with the atom through time-delayed feedback. In the ideal limit, the coupling points are located at $x_m=mx_0$, with $G_m=1$ and $\phi_m=0$. The generalized model is parametrized by the coupling-strength factors $G_m$, dimensionless positions $u_m=x_m/x_0$, and relative coupling phases $\phi_m$, where $\tau_0=x_0/v$ is the propagation time over the reference distance $x_0$. The three disorder channels modify the weights, propagation phases, and complex orientations of the interference paths, respectively. A common shift of all coupling phases is physically redundant. Therefore, only $N-1$ relative phases are independent.}
\label{fig:model-scheme}
\end{figure}

\subsection{Model Hamiltonian}
We formulate the model with arbitrary coupling positions and complex local couplings from the outset, so that coupling-strength, position, and coupling-phase imperfections can be treated as parameter subspaces of the same Hamiltonian in  Figure~\ref{fig:model-scheme}. The ideal configuration with equally spaced coupling points, equal coupling strengths, and zero relative phases is recovered as the special case $u_m = m$, $G_m = 1$, and $\phi_m = 0$ of this generalized model $(\hbar \equiv 1)$, where the Hamiltonian reads $\hat H=\hat H_a+\hat H_w+\hat H_{\mathrm{int}}$ with $\hat H_a=\omega_0\hat\sigma_+\hat\sigma_-$, $\hat H_w=\int_0^\infty dk\,\omega_k \hat a_k^\dagger \hat a_k$, $\hat H_{\mathrm{int}}=\sum_{m=1}^{N}\int_0^\infty dk\, (\mathrm{g}_{km}\hat a_k^\dagger\hat\sigma_-+\mathrm{g}_{km}^*\hat a_k\hat\sigma_+)$, $\hat{\sigma}_{-}=|g\rangle\langle e|$. The generalized local coupling amplitude is defined as $\mathrm{g}_{km}=\sqrt{\Gamma v / \pi}\,G_m e^{i\phi_m}\sin(kx_m)$ with $x_m=u_mx_0$. In this parametrization, $G_m$, $u_m=x_m/x_0>0$, and $\phi_m$ denote the dimensionless coupling-strength factor, the dimensionless coupling position, and the local coupling phase, respectively.
For the $s$th statistical disorder realization, we sample the three channels as
\begin{align}
G_m^{(s)} = 1+\eta_G r_{G,m}^{(s)},~ u_m^{(s)} = m+\eta_x r_{x,m}^{(s)},~ \widetilde{\phi}_m^{(s)} = \eta_\phi r_{\phi,m}^{(s)},
\label{eq:disorder-realizations}
\end{align}where $r_{G,m}^{(s)}$, $r_{x,m}^{(s)}$, and $r_{\phi,m}^{(s)}$ denote the random variables for the coupling strength, position, and phase at the $m\text{th}$ coupling point in the $s\text{th}$ realization, respectively, which are mutually independent and uniformly distributed over the interval $[-1, 1]$ (denoted by $U[-1, 1]$). The dimensionless parameters $\eta_G$, $\eta_x$, and $\eta_\phi$ set the maximum strength, position, and raw phase deviations, respectively, so $\widetilde{\phi}_m^{(s)}\in[-\eta_\phi,\eta_\phi]$. The common phase is removed by defining
$\phi_m^{(s)}=\widetilde{\phi}_m^{(s)}
-\widetilde{\phi}_1^{(s)}$, which fixes $\phi_1=0$ without changing
any physical phase difference. The gauge-fixed relative phases can lie in $[-2\eta_\phi,2\eta_\phi]$ and are not statistically independent after this subtraction, although the raw offsets are. We retain only realizations satisfying $G_m\geq0$ and $0<u_1<\cdots<u_N$. In formulas for a specified realization, the superscript $(s)$ is omitted for readability. The ideal configuration is recovered when $\eta_G=\eta_x=\eta_\phi=0$. The perfect mirror at $x=0$ selects sine-like standing-wave modes. Within the wide-band delay model, the waveguide dispersion is linearized around the relevant transition frequency as $\omega_k=vk$, and the local coupling envelope is taken to be frequency independent. The parameter $\Gamma$ denotes the spontaneous-emission rate of a single coupling point into the corresponding infinite-waveguide continuum in the wide-band limit under the normalization adopted above. The frequency $\omega_0$ is understood as the renormalized atomic transition frequency after the principal-value frequency shift has been absorbed. The extension of the positive-frequency integral used to obtain the retarded delay kernel, together with the limitations of the wide-band approximation, is discussed in  Appendix~\ref{app:delay-equation}.

\subsection{Equation of motion and solution}

For an initially excited atom and an empty waveguide, excitation-number conservation restricts the state to the single-excitation manifold, parametrized as $|\psi(t)\rangle  = \varepsilon(t)|e,0\rangle + \int_0^\infty dk\,\varphi(k,t)\hat a_k^\dagger|\mathrm{g},0\rangle$ with initial probability amplitudes $\varepsilon(0)=1$ and $\varphi(k,0)=0$.
We use $\varphi(k,t)$ and $\varphi(x,t)$ consistently for the photon amplitudes in momentum and real space, respectively. The real-space transformation is chosen to preserve the one-photon normalization, $\int_{0}^{\infty}dx\,|\varphi(x,t)|^{2}=\int_{0}^{\infty}dk\,|\varphi(k,t)|^{2}$. Consequently, $\varphi(x,t)$ has dimensions of $L^{-1/2}$, and the dimensionless photon-density quantity plotted below is $x_0|\varphi(x,t)|^{2}$. Projecting the Schr\"odinger equation onto the atomic and one-photon sectors yields 
\begingroup
\begin{align}
  \dot{\varepsilon}(t)
  =&-\imag\omega_0\varepsilon(t)
  \notag\\
  &-\imag\sqrt{\frac{\Gamma v}{\pi}}\sum_{m=1}^{N}G_m\e^{-\imag\phi_m}\int_0^\infty dk\,\sin(kx_m)\varphi(k,t),
  \label{eq:eom-eps}\\[1.5ex]
  \dot{\varphi}(k,t)
  =&-\imag\omega_k\varphi(k,t)
  \notag\\
  &\rule{0pt}{5.5ex}-\imag\sqrt{\frac{\Gamma v}{\pi}}\sum_{m=1}^{N}G_m\e^{\imag\phi_m}\sin(kx_m)\varepsilon(t).
  \label{eq:eom-phi}
\end{align}
\endgroup
Integrating out the waveguide modes gives the generalized delay equation
\begin{widetext}
\begin{equation}
\begin{aligned}
\dot{\varepsilon}(t)
&= -\imag\omega_0\varepsilon(t)
-\frac{\Gamma}{2}\sum_{m,n=1}^{N}
G_mG_ne^{\imag(\phi_n-\phi_m)}
\varepsilon[t-|u_m-u_n|\tau_0]
\Theta[t-|u_m-u_n|\tau_0] \\
&\quad +\frac{\Gamma}{2}\sum_{m,n=1}^{N}
G_mG_ne^{\imag(\phi_n-\phi_m)}
\varepsilon[t-(u_m+u_n)\tau_0]
\Theta[t-(u_m+u_n)\tau_0],
\end{aligned}
\label{eq:delay-equation-general}
\end{equation}
\end{widetext}
where the first delayed term represents direct propagation between
two coupling points, whereas the second accounts for emission
from one point, reflection by the mirror, and reabsorption at
another point. Their delays are determined by
$|u_m-u_n|\tau_0$ and $(u_m+u_n)\tau_0$, respectively.
The detailed derivation of Eq.~\eqref{eq:delay-equation-general}
is given in Appendix~\ref{app:delay-equation}. Their weights and relative phases are explicitly controlled by $G_mG_n\e^{\imag(\phi_n-\phi_m)}$, while their delays are set by $u_m$ and $u_n$.  The opposite signs of the two memory terms originate from sine-like standing-wave mode selected by the mirror.  The ideal delay equation follows by setting $G_m=1$, $u_m=m$, and $\phi_m=0$. Fourier transforming the solution for waveguide-mode amplitude into coordinate space gives
\begin{widetext}
\begin{equation}
\varphi(x,t) = -\imag\sqrt{\frac{\Gamma}{2v}} \sum_{m=1}^{N} G_m e^{\imag\phi_m} \left\{ \varepsilon[t-|\tau-u_m\tau_0|]\Theta[t-|\tau-u_m\tau_0|] - \varepsilon[t-(\tau+u_m\tau_0)]\Theta[t-(\tau+u_m\tau_0)] \right\},
\label{eq:real-space-field}
\end{equation}
\end{widetext}
where $\tau=x/v$. The first contribution in Eq.~\eqref{eq:real-space-field} is the directly propagating field, whereas the second is the mirror-reflected contribution. The factor $G_m\e^{\imag\phi_m}$ appears because the photon equation of motion contains the emission amplitude $\mathrm{g}_{km}$. The ideal field is recovered for $G_m=1$, $u_m=m$, and $\phi_m=0$. The derivation of Eq.~\eqref{eq:real-space-field}, together with
the bound-pole decomposition, normalization, and spatial support
of its nondecaying component, is given in Appendix~\ref{app:field-expression}.
The Laplace transform converts each retarded shift in Eq.~\eqref{eq:delay-equation-general} into an exponential factor, and the transformed excitation amplitude satisfies
\begin{widetext}
\begin{equation}
\tilde{\varepsilon}(s)
=\frac{\varepsilon(0)}{s+\imag\omega_0+
\frac{\Gamma}{2}\sum_{m,n=1}^{N}
G_mG_n\e^{\imag(\phi_n-\phi_m)}
\left[\e^{-s|u_m-u_n|\tau_0}-\e^{-s(u_m+u_n)\tau_0}\right]}.
  \label{eq:laplace-solution}
\end{equation}
The residue representation below assumes that all poles retained in the inverse Laplace transform are isolated and simple, such that $D(z_j)=0$ and $D'(z_j)\neq0$. Because the delay equation has a transcendental characteristic function, it possesses infinitely many complex roots. The numerical pole sum therefore includes all roots inside a specified search contour and needs to be checked for convergence as that contour is enlarged.
Inverting Eq.~\eqref{eq:laplace-solution} by summing the residues at its poles $s_k$ expands the excitation amplitude as

\begin{equation}
\varepsilon(t)=\sum_k
\frac{\varepsilon(0)\e^{s_kt}}
{1+\frac{\Gamma}{2}\sum_{m,n=1}^{N}
G_mG_n\e^{\imag(\phi_n-\phi_m)}
\left[-|u_m-u_n|\tau_0\e^{-s_k|u_m-u_n|\tau_0}
+(u_m+u_n)\tau_0\e^{-s_k(u_m+u_n)\tau_0}\right]}.
  \label{eq:residue-sum}
\end{equation}
The poles $s_k$, and hence the decay rates and oscillation frequencies, are the zeros of the transformed denominator
\begin{equation}
s_k+\imag\omega_0+
\frac{\Gamma}{2}\sum_{m,n=1}^{N}
G_mG_n\e^{\imag(\phi_n-\phi_m)}
[\e^{-s_k|u_m-u_n|\tau_0}
-\e^{-s_k(u_m+u_n)\tau_0}]=0.
  \label{eq:pole-dimensional}
\end{equation}
\end{widetext}
Equations~\eqref{eq:disorder-realizations}-\eqref{eq:pole-dimensional} form the complete generalized chain used in this work: the Hamiltonian, disorder realizations, state expansion, amplitude equations, delay equation, real-space field, Laplace solution, residue expansion, and pole equation all retain the explicit dependence on $G_m$, $u_m$, and $\phi_m$. They directly link the microscopic disorder realization to the experimentally observable long-time atomic and field dynamics. The pole with $\mathrm{Re}\,s_k<0$ yields a decaying contribution, whereas a pole with $s_k=-\imag\omega_k$ remains on the imaginary axis and yields a perfect bound-state contribution.

\subsection{Discussion of generalized bound-state conditions}
\label{sec:general-bound-state-conditions}
\label{sec:common-matched-root-diagnostics}

In this subsection, we derive the analytical conditions for the formation
of exact bound states in the generalized semi-infinite waveguide model
and determine when one, two, or three distinct bound-state roots can
coexist for the same physical parameter set. The bound state is associated with a pole of the excitation
amplitude on the imaginary axis. We therefore seek purely
imaginary solutions of Eq.~\eqref{eq:pole-dimensional},
writing $s_j=-\imag\omega_j$ with
$\omega_j\in\mathbb{R}$.
Multiplying Eq.~\eqref{eq:pole-dimensional} by the
propagation time $\tau_0$ and defining the dimensionless
pole variable $z=s\tau_0$, the coupling parameter
$\gamma=\Gamma\tau_0$, and the atomic frequency
$\Delta=\omega_0\tau_0$, we obtain
\begin{equation}
\begin{aligned}
D(z)=&z+\imag\Delta
+\frac{\gamma}{2}\sum_{m,n=1}^{N}
G_mG_n\e^{\imag(\phi_n-\phi_m)}
\\
&\times
[
\e^{-z|u_m-u_n|}
-\e^{-z(u_m+u_n)}
].
\end{aligned}
\label{eq:pole-general}
\end{equation}
For such an imaginary-axis pole, we introduce the dimensionless
angular frequency $\vartheta_j=\omega_j\tau_0$, so that
$z_j=s_j\tau_0=-i\vartheta_j$.
Evaluating Eq.~\eqref{eq:pole-general} at a general real angular-frequency variable $\theta$ motivates the definitions
\begin{align}
B(\theta)
&=
\sum_{m=1}^{N}
G_m\e^{\imag\phi_m}\sin(u_m\theta),
\label{eq:general-dark-revised}
\\
S(\theta)
&=
\frac{1}{2}\sum_{m,n=1}^{N}G_mG_n
\{
\sin[\phi_n-\phi_m+\theta|u_m-u_n|]
\notag\\
&\hspace{1.25cm}
-\sin[\phi_n-\phi_m+\theta(u_m+u_n)]
\},
\label{eq:general-matching-full}
\\
F(\theta)
&=
\Delta-\theta+\gamma S(\theta).
\label{eq:general-match-revised}
\end{align}
Using the trigonometric identity
$\cos[(u_m-u_n)\theta]-\cos[(u_m+u_n)\theta]
=2\sin(u_m\theta)\sin(u_n\theta)$, the imaginary-axis
pole function separates as
\begin{equation}
D(-\imag\theta)
=
\gamma|B(\theta)|^2+\imag F(\theta).
\label{eq:radiation-dispersion}
\end{equation}
Because $\gamma|B(\theta)|^2\geq0$, an exact
imaginary-axis pole necessarily satisfies
\begin{equation}
B(\theta_j)=0,
\quad
F(\theta_j)=0.
\label{eq:general-bound-condition}
\end{equation}
For complex local couplings, the condition
$B(\theta_j)=0$ is equivalent to the two real conditions
$\operatorname{Re}B(\theta_j)=0$ and
$\operatorname{Im}B(\theta_j)=0$.
For an isolated simple pole, the residue denominator is
\begin{equation}
\begin{aligned}
\mathcal N(\theta_j)
=&D'(z)\big|_{z=-\imag\theta_j}
\\
=&1+\frac{\gamma}{2}\sum_{m,n=1}^{N}
G_mG_n\e^{\imag(\phi_n-\phi_m)}
\\
&\times
[
-|u_m-u_n|
\e^{\imag\theta_j|u_m-u_n|}
\\
&\hspace{0.65cm}
+(u_m+u_n)
\e^{\imag\theta_j(u_m+u_n)}
].
\end{aligned}
\label{eq:residue-denominator-general}
\end{equation}
Thus, a zero $\theta_j$ of $B(\theta)$ is only a dark root. It becomes a physical bound-state root only if it satisfies $F(\theta_j) = 0$, remains simple, is normalizable, and satisfies $\mathcal{N}(\theta_j) \neq 0$. To classify the possible bound-state responses, we define
\begin{equation}
\begin{aligned}
\mathcal R_{\mathrm B}(\gamma,\Delta)
=
\{
\theta\in\mathbb R:\,
&B(\theta)=0,\,
\Delta=\theta-\gamma S(\theta),
\\
&\mathcal N(\theta)\neq0
\}.
\end{aligned}
\label{eq:general-bound-root-set}
\end{equation}
For a fixed disorder realization and fixed values of
$(\gamma,\Delta)$, the set
$\mathcal R_{\mathrm B}(\gamma,\Delta)$ contains all admissible
real-frequency roots that satisfy the radiation-cancellation,
frequency-matching, and simple-pole conditions. The associated
real-space field is also normalizable. We denote the
cardinality of this set by
$M=\operatorname{card}\mathcal R_{\mathrm B}$. Thus, $M$ is the
number of exact simple bound-state roots: $M=1$, $M=2$, and
$M=3$ correspond to the one-, two-, and three-root cases,
respectively.

\noindent{(i) One-root bound state.}
For $M=1$, the set $\mathcal R_{\mathrm B}$ contains a single
real dark root $\theta_1$ satisfying $B(\theta_1)=0$. For a
given coupling parameter $\gamma$, the matching condition
fixes the corresponding dimensionless atomic frequency
$\Delta$. {(ii) Two-root bound state.}
For $M=2$, the set $\mathcal R_{\mathrm B}$ contains two
distinct real dark roots $\theta_1$ and $\theta_2$ satisfying
$B(\theta_1)=B(\theta_2)=0$. The exact two-root response
requires the two root-matching conditions to yield the same
positive value of $\gamma$ and one common value of $\Delta$. {(iii) Three-root bound state.}
For $M=3$, the set $\mathcal R_{\mathrm B}$ contains three
distinct real dark roots $\theta_1$, $\theta_2$, and
$\theta_3$ satisfying
$B(\theta_1)=B(\theta_2)=B(\theta_3)=0$. These roots coexist
for the same physical parameters only when the pairwise
matching conditions for two independent root pairs yield the
same positive value of $\gamma$ and all three roots give the
same value of $\Delta$. Using the dimensionless time $T=t/\tau_0$, the bound-state
contribution associated with an $M$-root set is
\begin{equation}
\varepsilon_{\mathrm B}^{(M)}(T)
=
\sum_{j=1}^{M}
Z_j\e^{-\imag\theta_jT},
\quad
Z_j=\frac{1}{\mathcal N(\theta_j)}.
\label{eq:common-bound-response}
\end{equation}
Consequently, one root produces a time-independent nondecaying population, two roots can give rise to one normalized angular beat frequency $|\theta_2 - \theta_1|/\gamma$, and three roots produce up to three pairwise normalized angular beat frequencies $|\theta_j - \theta_l|/\gamma$. The same bound-state roots and residue factors determine the nondecaying real-space photonic field, which strictly vanishes outside the outermost coupling point, highlighting the localized nature of these states. The derivations of the two- and three-root compatibility conditions, the pole decomposition, the spatial mode profiles, and the normalization of each simple bound state are provided in Appendices~\ref{app:general-bound-state-conditions} and ~\ref{app:field-expression}.

In the clean limit, $G_m=1$, $u_m=m$, and $\phi_m=0$, the
generalized conditions reduce to the analytical clean-system
root families derived in
Appendix~\ref{app:general-bound-state-conditions}. This limit
is used only to construct the clean reference configurations
from which the poles are continued under imperfections. The
numerical parameters of these reference configurations are
specified together with the corresponding calculations.

\subsection{Scope of the imperfection analysis}

We distinguish realization-dependent compensation from passive statistical robustness.
Under realization-dependent compensation, a given imperfection realization is treated as known, and the available control parameters are retuned until the radiation-cancellation and dispersive-matching conditions are simultaneously satisfied. In passive
robustness, no realization-dependent compensation is
allowed. The exact stre\-ngth- and position-imperfect solutions below concern exact bound-state recovery through realization-dependent compensation. The phase-imperfect optimization retains finite linewidths and is
therefore interpreted as suppression of decay rather than
exact restoration. The ensemble calculations quantify the success probability of the realization-dependent compensation procedure but do not establish passive statistical robustness against arbitrary disorder.

\subsection{Finite-linewidth and beat-frequency signatures of bound and quasibound components}
\label{sec:finite-linewidth-diagnostics}

For an imperfect configuration, a pole continued from a clean bound
root generally takes the form
$z_j=-\kappa_j-i\vartheta_j$, where
$\kappa_j=-\operatorname{Re}z_j$ is the dimensionless linewidth and
$\vartheta_j=-\operatorname{Im}z_j$ is the dimensionless angular
frequency. An exact bound pole has $\kappa_j=0$, whereas
$\kappa_j>0$ identifies a finite-lifetime pole. With
$\Gamma t=\gamma T$, the pole-predicted normalized angular beat
frequency is
\begin{equation}
\frac{\Omega_{jl}^{(\mathrm{pole})}}{\Gamma}
=
\frac{
\left|\operatorname{Im}z_j-\operatorname{Im}z_l\right|
}{\gamma}.
\label{eq:pole_angular_frequency}
\end{equation}
To extract the corresponding normalized angular frequency from the numerical dynamics, we sample the atomic population $P_e(\Gamma t) = |\varepsilon(t)|^2$ on a uniform time grid. After removing a linear trend and applying a Hann window, we compute its fast Fourier transform (FFT), which initially returns the frequency $f_{\mathrm{FFT}}$ in cycles per unit $\Gamma t$. Before comparing it with the pole prediction, we convert it to the normalized angular frequency. The corresponding conversion and relative angular-frequency error are

\begin{equation}
\begin{aligned}
\frac{\Omega_{\mathrm{FFT}}}{\Gamma}
&=2\pi f_{\mathrm{FFT}},
\\[3pt]
E_{\Omega,jl}
&=
\frac{
|
\Omega_{\mathrm{FFT}}/\Gamma
-
\Omega_{jl}^{(\mathrm{pole})}/\Gamma
|
}{
\Omega_{jl}^{(\mathrm{pole})}/\Gamma
},
\\[3pt]
E_{\Omega,jl}^{(\%)}
&=100E_{\Omega,jl},
\end{aligned}
\label{eq:spectral-diagnostics}
\end{equation}
where $f_{\mathrm{FFT}}$ is obtained from the linearly detrended and
Hann-windowed population calculated from the generalized delay
equation, whereas the converted quantity
$\Omega_{\mathrm{FFT}}/\Gamma=2\pi f_{\mathrm{FFT}}$ is used in all
spectral plots and numerical comparisons. The resolved angular-frequency
peak alone does not establish the bound state. The corresponding poles
are additionally required to have zero linewidth and satisfy the exact
radiation-cancellation and frequency-matching conditions in
Eq.~\eqref{eq:general-bound-condition}, together with the simple-pole and normalizability
requirements in Eq.~\eqref{eq:general-bound-root-set}.

\section{Bound-state restoration after coupling-strength imperfections}
\label{sec:strength-disorder}

For coupling-strength imperfections, the coupling positions and phases remain fixed at
$u_m=m$ and $\phi_m=0$, whereas the real coupling-strength factors are sampled as
$G_m=1+\eta_G r_{G,m}$ according to Eq.~\eqref{eq:disorder-realizations}.
The generalized radiation-cancellation and dispersive-matching conditions in Eq.~\eqref{eq:general-bound-condition} then reduce to
\begin{align}
B_G(\theta)
&=\sum_{m=1}^{N}G_m\sin(m\theta)=0,
\label{eq:strength-root-test}\\
\Delta
&=\theta-\gamma S_G(\theta),
\label{eq:strength-matching-condition}
\end{align}
where
\begin{equation}
S_G(\theta)
=\sum_{m>n}G_mG_n\sin[(m-n)\theta].
\label{eq:strength-dispersive-shift}
\end{equation}
The exact bound root needs to satisfy Eqs.~\eqref{eq:strength-root-test}
and \eqref{eq:strength-matching-condition} at the same real
frequency and remains a simple pole,
$D_G'(-\imag\theta_j)\neq0$. Coupling-strength imperfections
leave the propagation phases unchanged but modify the weights
of the interfering emission and mirror-reflected paths.

\subsection{Loss of the clean bound roots at fixed parameters}

\begin{figure*}[!t]
\centering
\includegraphics[width=2.085\columnwidth]{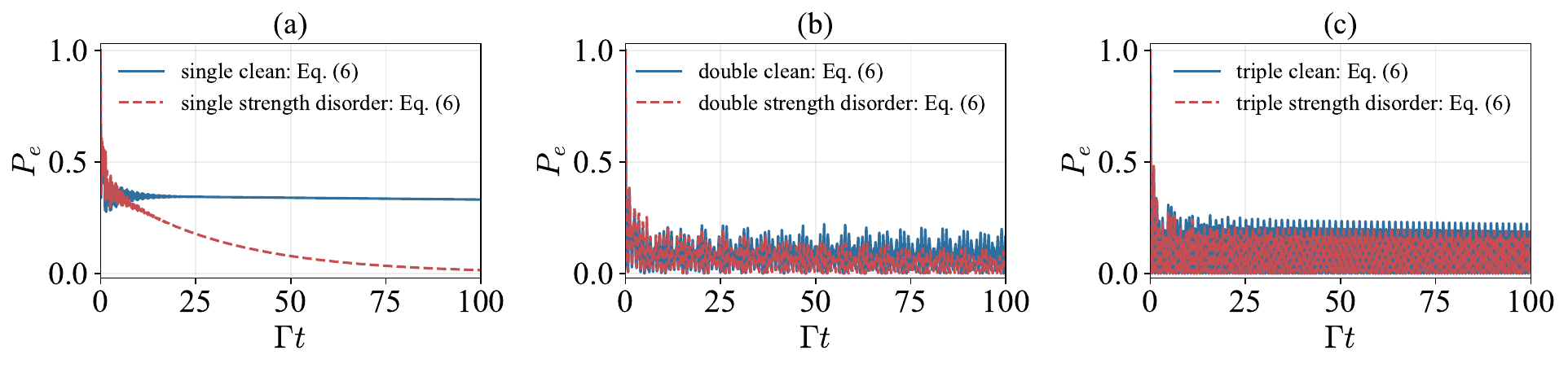}
\caption{For the seed-11 coupling-strength-disorder realization, with the
clean-system parameters unchanged, the atomic excitation probability
$P_e(t)=|\varepsilon(t)|^2$ is obtained from  Eq.~\eqref{eq:delay-equation-general}. From left to
right, (a)--(c) target the clean single-, double-, and
triple-root configurations. Blue solid curves represent the clean systems, whereas red dashed
curves represent the disordered systems with the same clean parameter
sets $(\theta_1,\gamma,\Delta)
=(1.25\pi,0.05\pi,1.3329\pi)$,
$(\theta_1,\theta_2,\gamma,\Delta)
=(1.25\pi,2.5\pi,0.2209\pi,1.6161\pi)$, and
$(\theta_1,\theta_2,\theta_3,\gamma,\Delta)
=(1.5\pi,2\pi,2.5\pi,0.125\pi,2\pi)$,
respectively. The decaying or damped red curves show that these clean
parameters no longer simultaneously satisfy Eqs.~\eqref{eq:strength-root-test} and
\eqref{eq:strength-matching-condition}.}
\label{fig:strength-unmatched-population}
\end{figure*}

To determine whether the clean bound roots survive without parameter retuning, we
consider a representative realization with $N=7$, $\eta_G=0.1$, and pseudorandom seed
11. The sampled factors are $G_1=0.9257$, $G_2=0.9999$, $G_3=1.0203$,
$G_4=0.9057$, $G_5=0.9296$, $G_6=1.0856$, and $G_7=0.9141$.
The positions and coupling phases remain $u_m=m$ and $\phi_m=0$. We first introduce
these coupling-strength variations while retaining the clean roots and the corresponding
clean values of $\gamma$ and $\Delta$. Figure~\ref{fig:strength-unmatched-population} shows that changing the coupling weights while holding the clean control parameters fixed generally moves the target poles away from the imaginary axis. Consequently, the red curves represent decaying or finite-time quasibound responses rather than exact bound-state dynamics. To resolve the continued poles underlying the damped multi-root responses, we use the same seed-11 random profile at $\eta_G=0.2$ and retain the clean values of $\gamma$ and $\Delta$. For the double-state control, the continued linewidths are
$\vec{\kappa}^{\,(\mathrm{str})}=(2.2\times10^{-2},\,4.07\times10^{-4})$. The pole prediction $\Omega_{12}^{(\mathrm{pole})}/\Gamma=5.608$ agrees with the resolved FFT value $\Omega_{\mathrm{FFT}}/\Gamma=5.6275$, corresponding to $E_{\Omega,12}=3.6\times10^{-3}$ (0.36\%). For the triple-reference control, the $2\pi$ pole remains exact, whereas the two companion poles have linewidths $1.95\times10^{-4}$. Thus, $\vec{\kappa}^{\,(\mathrm{str})}=(1.95\times10^{-4},\,0,\,1.95\times10^{-4})$. The degenerate predictions $\Omega_{12}^{(\mathrm{pole})}/\Gamma =\Omega_{23}^{(\mathrm{pole})}/\Gamma=3.9142$ agree with the resolved FFT value
$\Omega_{\mathrm{FFT}}/\Gamma=3.9264$, giving an error of 0.31\%. The agreement between the predicted and resolved frequencies shows that the damped oscillations are governed by the continued pole pairs. However, their nonzero linewidths distinguish these responses from fully restored exact bound states. The exact $2\pi$ root in the triple reference is a special consequence of the unchanged integer coupling positions and $\Delta=2\pi$. However, it does not make the other two finite-linewidth poles exact.

\subsection{Exact bound-state recovery for one-, two-, and three-root configurations}
\begin{figure*}[!t]
\centering
\includegraphics[width=2.085\columnwidth]{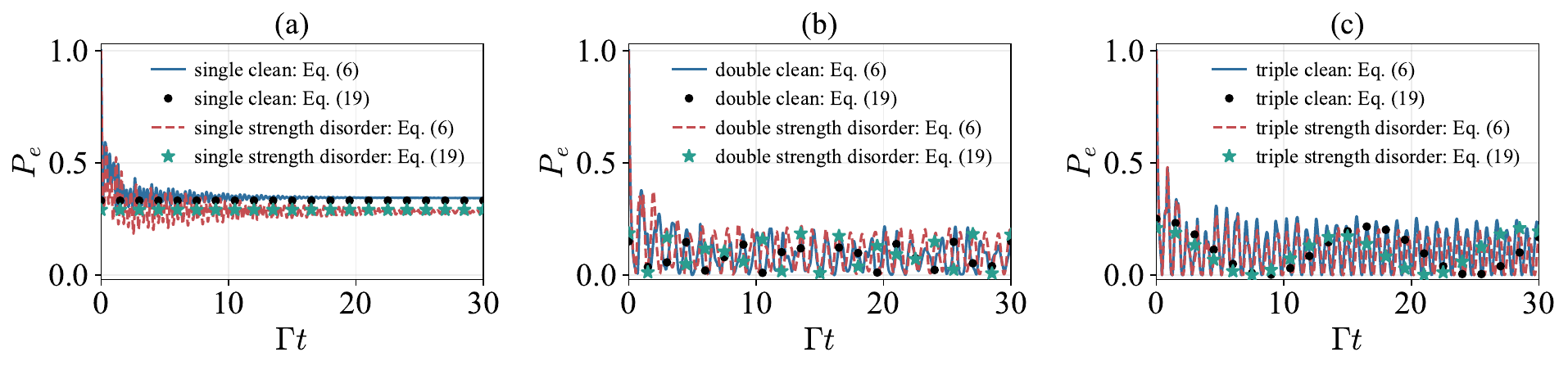}
\caption{Restored single-, double-, and triple-root excitation-population
dynamics for the seed-11 coupling-strength-disorder realization. (a)--(c) correspond to the single-, double-, and triple-root
configurations, respectively.
The blue solid curves show the clean-system atomic excitation
probability $P_e(t)=|\varepsilon(t)|^2$ obtained from Eq.~\eqref{eq:delay-equation-general}, whereas
the black dots show the bound-pole population
reconstructions obtained from Eq.~\eqref{eq:common-bound-response}.
The red dashed curves and teal stars show the results
from Eqs.~\eqref{eq:delay-equation-general} and \eqref{eq:common-bound-response}, respectively, for the
coupling-strength-disordered systems after the control parameters are
set to the matched values in the text.
Their late-time agreement confirms a stationary nondecaying
contribution in the single-root case and persistent beating in the
double- and triple-root cases.
The middle root of the triple-root configuration is fixed at
$\theta_2=2\pi$.}
\label{fig:strength-retuned-population}
\end{figure*}

\begin{figure*}[!t]
\centering
\includegraphics[width=2.085\columnwidth]{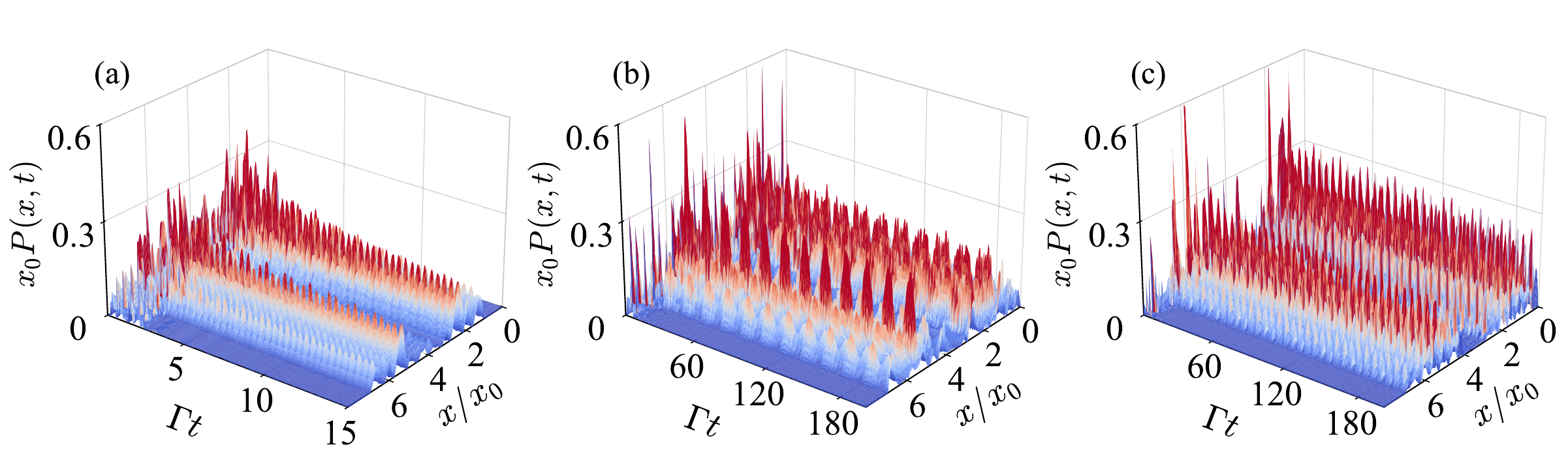}
\caption{Dimensionless real-space photon density $x_0P(x,t)$, where $P(x,t)=|\varphi(x,t)|^2$, for the retuned seed-11 coupling-strength realization with $N=7$ and $\eta_G=0.1$. (a)--(c) show the single-, double-, and triple-root solutions, respectively, calculated from Eq.~\eqref{eq:real-space-field} using the matched parameters specified in the text. The single-root field approaches a stationary confined profile, whereas the double- and triple-root fields exhibit persistent temporal modulation generated by interference among the corresponding exact bound poles.}
\label{fig:strength-retuned-field}
\end{figure*}

We next jointly solve Eqs.~\eqref{eq:strength-root-test} and
\eqref{eq:strength-matching-condition} for the specified seed-11
realization and then set the available control parameters to the
resulting matched values for this disordered realization. The single-root solution is $\theta_1=1.4045\pi$,
$\gamma=0.05\pi$, and $\Delta=1.5313\pi$.
The double-root solution is
$(\theta_1,\theta_2)=(1.4045\pi,2.4951\pi)$,
$\gamma=0.1676\pi$, and $\Delta=1.8295\pi$.
The triple-root solution is
$(\theta_1,\theta_2,\theta_3)
=(1.5049\pi,2\pi,2.4951\pi)$,
$\gamma=0.1247\pi$, and $\Delta=2\pi$.
The values above are displayed to four decimal places. The
corresponding maximum pole residuals, evaluated using the
full-precision numerical solutions, are $8.16\times10^{-16}$,
$1.38\times10^{-14}$, and $1.56\times10^{-8}$, respectively.
The nonzero values of $D_G'(-i\theta_j)$ confirm that the selected
poles are simple. Figure~\ref{fig:strength-retuned-population} verifies that simultaneous recovery of the radiat\-ion-cancellation and dispersive-matching conditions returns the selected poles to the imaginary axis. The single-root solution produces a stationary nondecaying population, whereas interference between two or three exact bound poles produces persistent population oscillations. The field distributions in Figure~\ref{fig:strength-retuned-field} provide the real-space
counterpart of the population dynamics in Figure~\ref{fig:strength-retuned-population}.
Coupling-stren\-gth errors do not change the coupling positions or propagation delays, but
they reweight the direct and mirror-reflected paths and consequently modify the dark
roots, dispersive shifts, pole residues, and internal field profiles. Exact restoration
therefore requires $B_G(\theta_j)=0$ and
$\Delta=\theta_j-\gamma S_G(\theta_j)$ to hold simultaneously for every selected root.
This constitutes exact bound-state recovery through realization-dependent compensation for a specified imperfection realization, rather than generic robustness against uncontrolled coupling-strength disorder.
\section{Bound-state restoration after position imperfections}
\label{sec:position-disorder}

The position-imperfection slice is obtained from Eq.~\eqref{eq:pole-general}
by setting $G_m=1$ and $\phi_m=0$, while using
$u_m=m+\xi_m$, where $\xi_m=\eta_x r_{x,m}$ and
$0<u_1<\cdots<u_N$. The radiation-cancellation and
dispersive-matching conditions then become
\begin{align}
B_x(\theta)
&=\sum_{m=1}^{N}\sin(u_m\theta)=0,
\label{eq:position-root-test}\\
\Delta
&=\theta-\gamma S_x(\theta),
\label{eq:position-matching-condition}
\end{align}
where
\begin{equation}
S_x(\theta)
=\sum_{m>n}\sin[(u_m-u_n)\theta].
\label{eq:position-dispersive-shift}
\end{equation}
The exact position-disordered bound root satisfies
Eqs.~\eqref{eq:position-root-test} and
\eqref{eq:position-matching-condition} at the same real frequency
and remains a simple pole, $D_x'(-\imag\theta_j)\neq0$.
Position variations modify both the propagation arguments in the
radiation amplitude and the direct and mirror-reflected delays.
The same displaced coordinates should therefore be used in the
radiation-cancellation condition, dispersive-matching condition, pole residue, and real-space field.
\subsection{Loss of the clean bound roots at fixed parameters}

\begin{figure*}[!t]
\centering
\includegraphics[width=2.085\columnwidth]{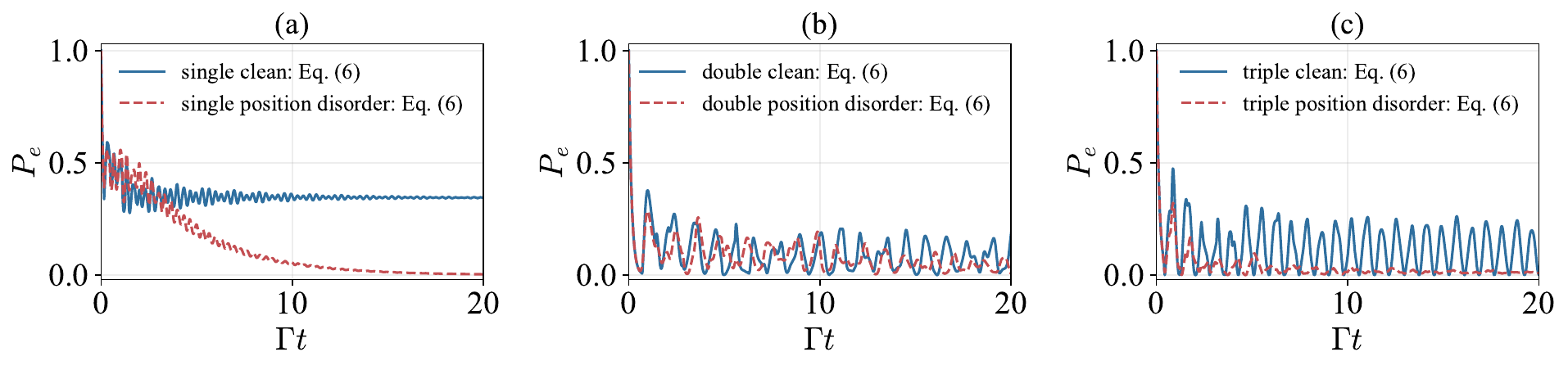}
\caption{For the seed-7 position realization introduced without
readjusting the clean-system parameters, the atomic excitation
probability $P_e(t)=|\varepsilon(t)|^2$ is obtained by solving
Eq.~\eqref{eq:delay-equation-general}. (a)--(c) correspond to the single-, double-, and
triple-root configurations, respectively. Blue solid curves represent
the clean systems, whereas red dashed curves represent the
position-disordered systems. From left to right, the clean parameter sets are
$(\theta_1;\gamma,\Delta)=(1.25\pi;0.05\pi,1.3328\pi)$,
$(\theta_1,\theta_2;\gamma,\Delta)
=(1.25\pi,2.5\pi;0.221\pi,1.6161\pi)$, and
$(\theta_1,\theta_2,\theta_3;\gamma,\Delta)
=(1.5\pi,2\pi,2.5\pi;0.125\pi,2\pi)$,
respectively.
The displaced systems no longer satisfy both position-dependent
bound-state conditions. Consequently, the single-root population
decays, while the double- and triple-root oscillations become damped.
The fixed root $\theta=2\pi$ is no longer dark in the triple-root case.}
\label{fig:position-unmatched-population}
\end{figure*}

For the fixed-parameter control, we consider a representative
realization with $N=7$, $\eta_x=0.1$, and pseudorandom
\begin{figure}[!t]
\centering
 \includegraphics[width=1.01\columnwidth, height=6.9cm]{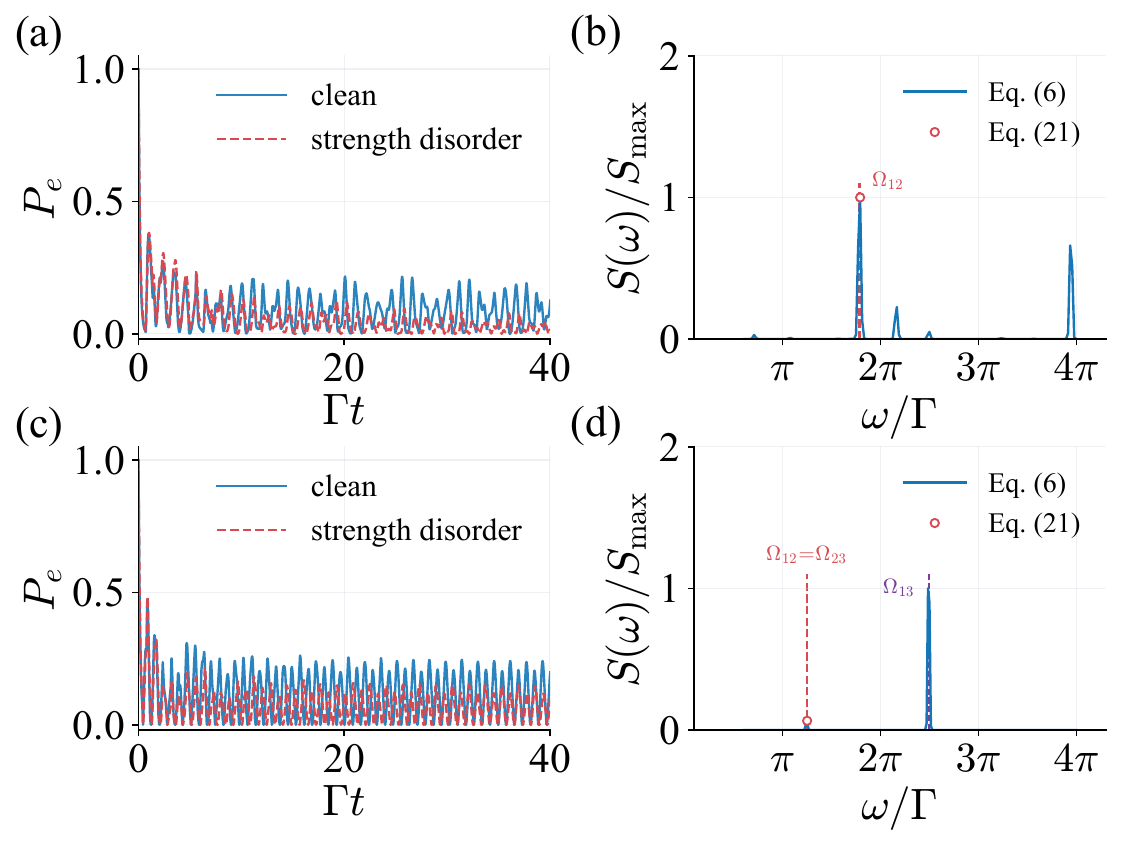}
\caption{Finite-linewidth beat diagnostics for coupling-strength disorder with
$N=7$, $\eta_G=0.2$, and the seed-11 random profile, while the clean
values of $\gamma$ and $\Delta$ are retained. (a) and (c)
compare the clean populations obtained from Eq.~\eqref{eq:delay-equation-general}, shown by the blue
solid curves, with the strength-disordered populations, shown by the
red dashed curves, for the double-state and triple-reference settings,
respectively. (b) and (d) show the corresponding normalized
spectral power $S(\omega)/S_{\max}$ of the disordered responses over
$\Gamma t\in[4,100]$ and $[4,140]$, respectively, plotted as a
function of the normalized angular frequency $\omega/\Gamma$. Each
spectrum is normalized by its maximum over the displayed frequency
interval. In the spectral panels, the vertical dashed lines indicate
the pairwise pole predictions
$\Omega_{jl}^{(\mathrm{pole})}/\Gamma$ from Eq.~\eqref{eq:pole_angular_frequency}, whereas the
labeled open circles mark the resolved FFT peaks nearest to the
corresponding predictions, with
$\Omega_{\mathrm{FFT}}/\Gamma$ obtained from Eq.~\eqref{eq:spectral-diagnostics}. In the
triple-reference setting, the root $\theta_2=2\pi$ remains exact,
whereas its companion poles retain finite linewidths. These finite
linewidths demonstrate that a resolved angular-frequency peak alone
does not establish an exact multiroot bound state.}
\label{fig:strength-beat-diagnostics}
\end{figure}
seed 7. The sampled dimensionless positions are $u_1=1.025$,
$u_2=2.0794$, $u_3=3.0551$, $u_4=3.945$,
$u_5=4.96$, $u_6=6.0747$, and $u_7=6.901$.
All coupling-strength factors and coupling phases remain fixed at
$G_m=1$ and $\phi_m=0$, respectively. We first retain the clean
root sets and the corresponding clean values of $\gamma$ and
$\Delta$ after replacing $u_m=m$ by the displaced coordinates in
the generalized delay equation. Figure~\ref{fig:position-unmatched-population} shows that merely
applying the clean parameters to the displaced geometry does not
preserve the target imaginary-axis poles. The displaced coordinates shift the dark roots and change the dispersive contribution, so the two position-dependent conditions are no longer satisfied simultaneously. The red curves are therefore interpreted as decaying or damped transients rather than as exact
position-disordered bound states. We quantify this loss of root matching using the same seed-7
displacement profile at $\eta_x=0.2$, while retaining the clean
values of $\gamma$ and $\Delta$. For the double-state control, the
continued linewidths are
$\vec{\kappa}^{\,(\mathrm{pos})}
=(2.62\times10^{-2},\,2.67\times10^{-2})$. The pole prediction $\Omega_{12}^{(\mathrm{pole})}/\Gamma=4.5553$ agrees with the resolved FFT value $\Omega_{\mathrm{FFT}}/\Gamma=4.5805$, corresponding to $E_{\Omega,12}=5.5\times10^{-3}$ (0.55\%). For the triple-reference control, the linewidths increase to
$(6.01\times10^{-1},5.61\times10^{-2},1.85\times10^{-2})$.
The spectral peak selected near the least-damped pair is
$\Omega_{\mathrm{FFT}}/\Gamma=5.0812$, compared with the pole
prediction $\Omega_{12}^{(\mathrm{pole})}/\Gamma=3.2004$, giving
$E_{\Omega,12}=0.5877$ (58.77\%). The agreement in the double-state control shows that its damped oscillations are governed by the continued pole pair. By contrast, the large angular-frequency error of the triple-reference response prevents its finite-time spectrum from being assigned to the target three-root beat structure. Together with the loss of the $\theta=2\pi$ dark root, this provides a frequency-domain rejection of an exact position-disordered triple bound state.
\subsection{Exact recovery of one- and two-root targets and failure of the specified triple-root target}

We next explicitly reconstruct the fully matched root sets by imposing the radiation-cancellation and dispersive-matching conditions with the same displaced coordinates. For the single-root solution, the retuned parameters are $\theta_1=1.323\pi$,
$\gamma=0.05\pi$, and $\Delta=1.4184\pi$.
The double-root solution is o-

\begin{figure*}[!t]
\centering
\includegraphics[width=2.085\columnwidth]{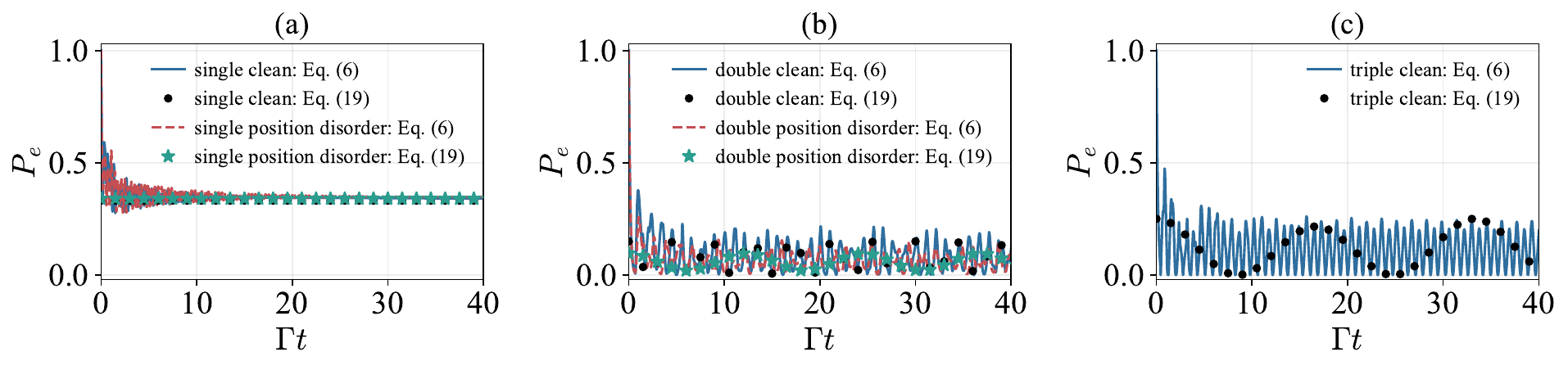}
\caption{Position-disorder population dynamics after root matching for
the seed-7 realization. (a)--(c) correspond to the single-,
double-, and triple-root configurations, respectively. The blue solid
and red dashed curves show the atomic excitation probability
$P_e(t)=|\varepsilon(t)|^2$ from Eq.~\eqref{eq:delay-equation-general} for the clean and
position-disordered systems, respectively, while the black dots and
teal stars denote the corresponding bound-pole reconstructions from Eq.~\eqref{eq:common-bound-response}. Retuning restores one simple root at
$\theta_1=1.3230\pi$ in (a) and two simple roots at
$(\theta_1,\theta_2)=(1.3230\pi,2.4478\pi)$ in (b),
producing a stationary nondecaying component and persistent beating,
respectively. In (c), only the clean triple-root reference is
shown because the position-disordered realization does not satisfy the
exact triple-root conditions. The required root $\theta=2\pi$ is no
longer dark after the coupling positions are displaced.}
\label{fig:position-retuned-population}
\end{figure*}
\begin{figure}[!t]
\centering
\includegraphics[width=\columnwidth]{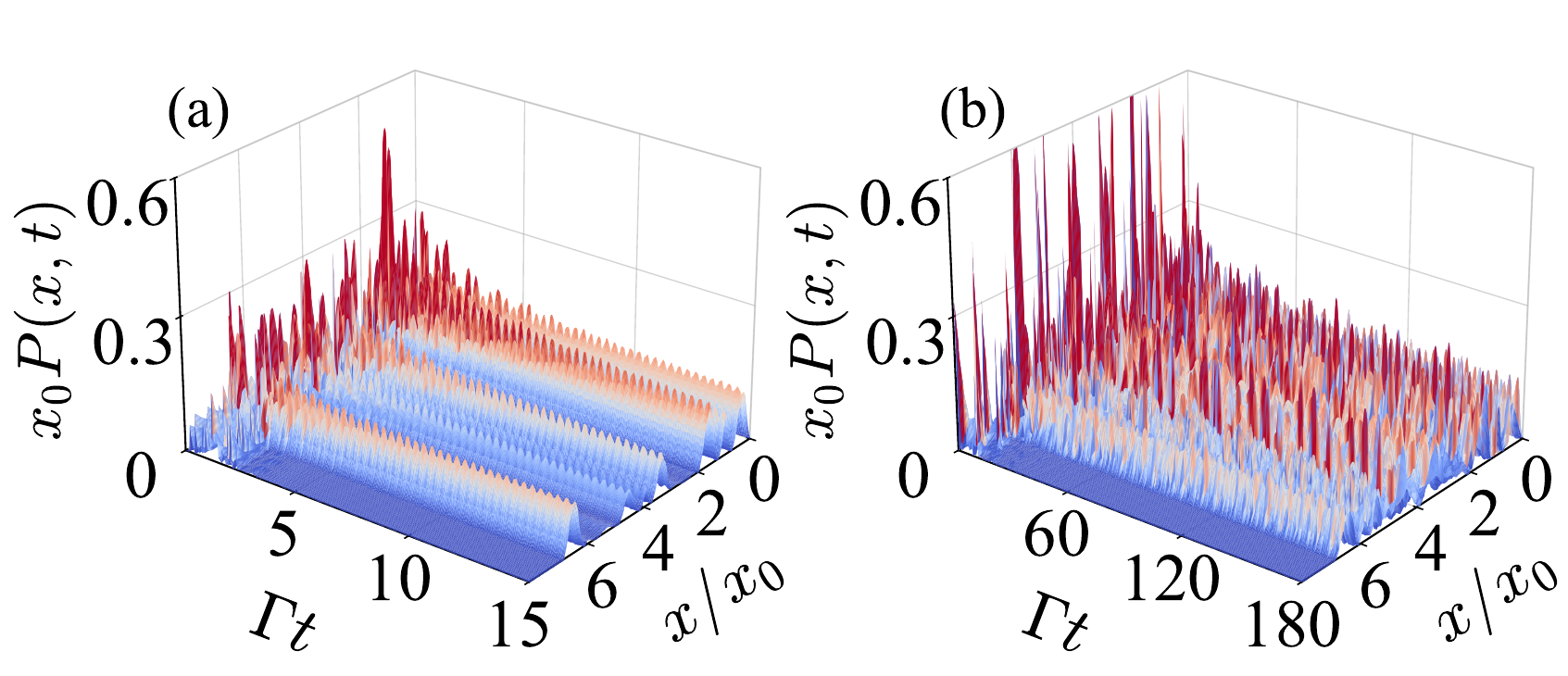}
\caption{Dimensionless real-space photon density $x_0P(x,t)$, where $P(x,t)=|\varphi(x,t)|^2$, for the retuned seed-7 position realization with $N=7$ and $\eta_x=0.1$. (a) and (b) show the single- and double-root solutions, respectively, calculated from Eq.~\eqref{eq:real-space-field} using the matched parameters specified in the text. The single-root field approaches a stationary confined profile, whereas the double-root field exhibits persistent temporal modulation generated by interference between the two exact bound poles. No triple-root field is shown because the displaced geometry does not satisfy $B_x(2\pi)=0$ and therefore does not generate a third bound-state residue.}
\label{fig:position-bound-field}
\end{figure}

\noindent btained at $(\theta_1,\theta_2)=(1.323\pi,2.4478\pi)$,
$\gamma=0.2393\pi$, and $\Delta=1.7759\pi$.
The values above are displayed to four decimal places. The
corresponding numerical residuals are
$\max_j|B_x(\theta_j)|\simeq5.71\times10^{-8}$ and
$\max_j|D_x(-\imag\theta_j)|\simeq2.19\times10^{-15}$
for the single-root solution, and $5.71\times10^{-8}$ and
$1.92\times10^{-14}$ for the double-root solution, respectively.
The nonzero derivative magnitudes,
$\min_j|D_x'(-\imag\theta_j)|\simeq1.7148$ and $4.4207$,
confirm that the selected poles are simple. As shown in Figure~\ref{fig:position-retuned-population}, the
long-time delay-equation dynamics agrees with the reconstructed
bound-pole contribution. One matched root produces a stationary
nondecaying population, whereas two matched roots produce a beat
frequency
$|\theta_2-\theta_1|=1.1248\pi$, corresponding to the period
$T_{\mathrm{beat}}^{(\Gamma t)}=1.3366$.
By contrast, the same position realization does not yield the
target triple-root configuration because the required
$\theta=2\pi$ root fails the radiation-cancellation condition.
Retuning $\gamma$ and $\Delta$ cannot repair a failure of
$B_x(\theta)=0$. Hence no position-disordered triple bound state
is asserted for this realization. The photon-density distributions provide a real-space characterization
of the same matched poles. Figure~\ref{fig:position-bound-field}
shows the fields obtained using the retuned single- and double-root
parameter sets in Figure~\ref{fig:position-retuned-population}. The
single-root solution approaches a stationary confined profile,
whereas the two-root solution produces a time-dependent interference
pattern.
Unlike coupling-strength imperfections, position variations change
the propagation phases themselves. The displaced coordinates enter
the radiation amplitude through $u_m\theta$ and the delay kernel
through both $|u_m-u_n|$ and $u_m+u_n$. They therefore modify the
dark roots, dispersive shifts, residues, and real-space field
profiles. Retuning the available parameters restores the single-
and double-root states when the two position-dependent conditions
can be satisfied simultaneously. It cannot restore the specified
triple-root target for this realization because changing $\gamma$
and $\Delta$ does not recover the failed radiative cancellation at
$\theta=2\pi$. These results illustrate realization-dependent compensation for a specified position realization, rather than generic robustness against uncontroll position disorder.

\section{Trivial bound roots and nontrivial quasibound responses
under coupling-phase imperfections }
\label{sec:phase-disorder}

The coupling-phase slice is obtained from Eq.~\eqref{eq:pole-general}
by setting $G_m=1$ and $u_m=m$, while allowing the relative phases
$\phi_m$ to vary under the gauge convention $\phi_1=0$. The
radiation-cancellation condition is then complex
\begin{equation}
B_\phi(\theta)
=\sum_{m=1}^{N}\e^{\imag\phi_m}\sin(m\theta)=0.
\label{eq:phase-root-test}
\end{equation}
The corresponding dispersive-matching condition is
\begin{equation}
\Delta=\theta-\gamma S_\phi(\theta),
\label{eq:phase-matching-condition}
\end{equation}
where
\begin{equation}
\begin{aligned}
S_\phi(\theta)
=\operatorname{Im} [
\frac{1}{2}\sum_{m,n=1}^{N}
\e^{\imag(\phi_n-\phi_m)}
(
\e^{\imag\theta|m-n|}
-\e^{\imag\theta(m+n)}
)
].
\end{aligned}
\label{eq:phase-dispersive-shift}
\end{equation}
\subsection{Trivial and nontrivial root conditions}
The phase-disordered roots are separated into trivial and
nontrivial classes. For every positive integer $q$, the equally
spaced geometry $u_m=m$ gives
\begin{equation}
B_\phi(q\pi)=0,
\quad
S_\phi(q\pi)=0,
\quad
\Delta=q\pi.
\label{eq:phase-trivial-root}
\end{equation}
\noindent The first equality follows from $\sin(mq\pi)=0$ for every coupling
point. The second follows because
$\e^{\imag q\pi|m-n|}=\e^{\imag q\pi(m+n)}$ for integer $m$ and
$n$. Consequently, the roots $\theta_q=q\pi$ remain dark
independently of the particular coupling phases. When
$\Delta=q\pi$ and $D_\phi'(-\imag q\pi)\neq0$, such a root is an
exact simple bound pole. We refer to these symmetry-enforced roots as trivial roots. Distinct trivial roots cannot by themselves form a double- or triple-root state for a fixed system because each one requires a different detuning $\Delta=q\pi$.
\begin{figure}[!t]
\centering
\includegraphics[width=1.01\columnwidth,height=6.9cm]{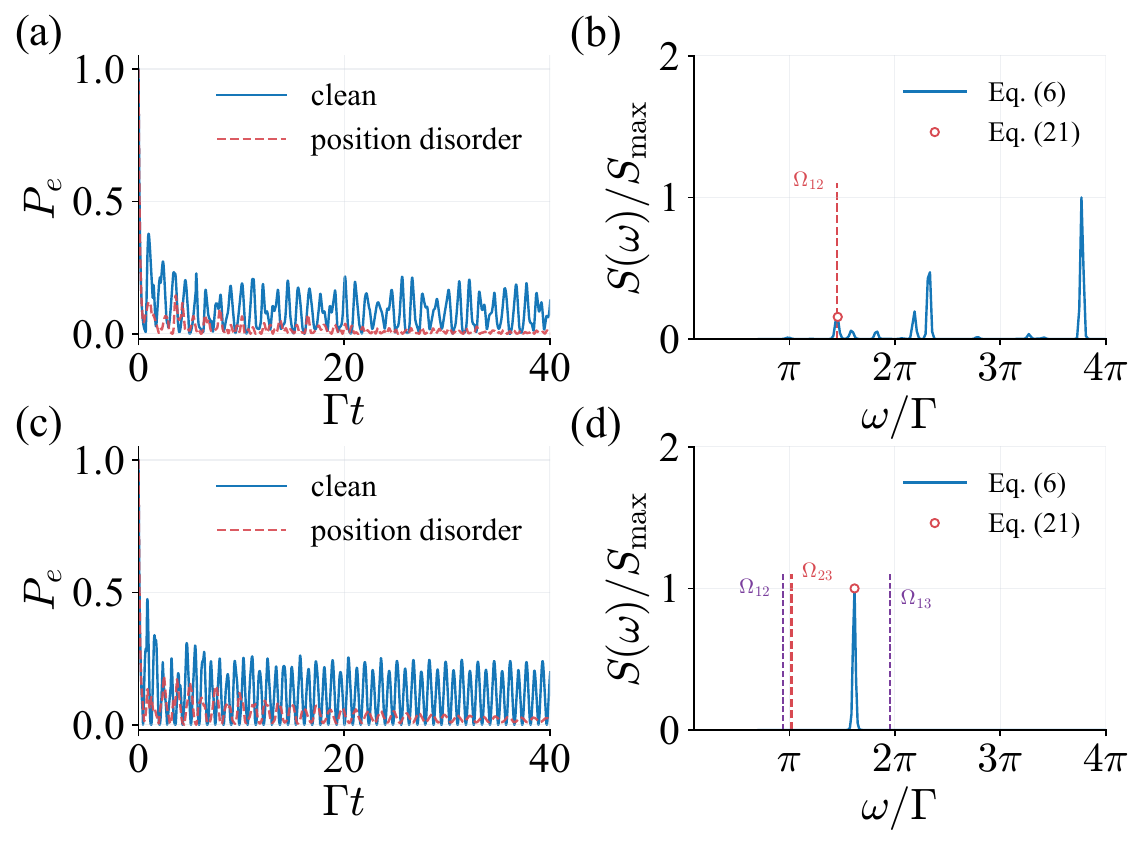}
\caption{Finite-linewidth beat characterization of position-disordered responses with
$N=7$, $\eta_x=0.2$, and the seed-7 displacement profile, while the
clean values of $\gamma$ and $\Delta$ are retained. (a) and (c)
compare the clean populations, shown by the blue solid curves, with
the position-disordered populations, shown by the red dashed curves,
for the double-state and triple-reference settings, respectively.
All population dynamics are obtained from Eq.~\eqref{eq:delay-equation-general}. In the clean
triple-reference setting, the root $\theta_2=2\pi$ is symmetry
protected, whereas position disorder breaks the corresponding
dark-root condition. (b) and (d) show the normalized spectral
power $S(\omega)/S_{\max}$ of the position-disordered responses over
$\Gamma t\in[4,100]$ and $[4,140]$, respectively. The spectra are
detrended and Hann windowed. The blue curves are Fourier transforms of
the dynamics obtained from Eq.~\eqref{eq:delay-equation-general}. Dashed lines indicate the pole-beat predictions
$\Omega_{jl}^{(\mathrm{pole})}/\Gamma$ from Eq.~\eqref{eq:pole_angular_frequency}, while open
circles mark the nearest resolved FFT peaks evaluated according to
Eq.~\eqref{eq:spectral-diagnostics}. The double-state peak agrees with the continued-pole
prediction, whereas the triple-reference response does not reproduce
the clean three-root structure.}
\label{fig:position-beat-diagnostics}
\end{figure}

\noindent
A multi-root configuration
containing one trivial root therefore requires one or more
nontrivial companion roots that satisfy the same global values of
$\gamma$ and $\Delta$. For a nontrivial frequency $\theta\notin\pi\mathbb{Z}$, the complex
condition $B_\phi(\theta)=0$ represents two independent real
conditions
\begin{equation}
\operatorname{Re}B_\phi(\theta)=0,
\quad
\operatorname{Im}B_\phi(\theta)=0.
\label{eq:phase-nontrivial-dark-condition}
\end{equation}
The real frequency supplies only one scalar variable for satisfying
these two equations. Moreover, neither $\gamma$ nor $\Delta$ enters $B_\phi(\theta)$. Retuning these two parameters can change the dispersive matching and the positions of the characteristic poles,
but it cannot generally close both radiation quadratures for a fixed, independently sampled disorder realization of the relative coupling phases. Exact nontrivial roots
therefore require additional phase correlations or independently
controllable coupling parameters.
\subsection{Exact trivial-root dynamics and multi-root obstruction}

We consider $N=7$, $\eta_\phi=0.1$, and pseudorandom seed~7. After removing the common phase by setting $\phi_1=0$, the gauge-fixed
relative phases are $\phi_1=0$, $\phi_2=0.0544$,
$\phi_3=0.0301$, $\phi_4=-0.0799$, $\phi_5=-0.065$,
$\phi_6=0.0497$, and $\phi_7=-0.124$. The last relative phase
can exceed $\eta_\phi$ in magnitude because the independently sampled
raw phases lie in $[-\eta_\phi,\eta_\phi]$ before the common reference
phase is subtracted. When trivial roots are included in the root search,
the phase-disordered single-root solution is obtained at
$\theta_1\simeq\pi$, with $\gamma=0.05\pi$ and $\Delta\simeq\pi$.
The numerical residuals are $|B_\phi(\theta_1)|\simeq6.49\times10^{-11}$
and $|D_\phi(-i\theta_1)|\simeq4.36\times10^{-17}$, while
$|D_\phi'(-i\theta_1)|\simeq1.6531$. This solution therefore contains an exact bound state associated with a simple imaginary-axis pole at the trivial root $\theta_1=\pi$. Figure~\ref{fig:phase-trivial-root-test} demonstrates that phase disorder does not eliminate
the symmetry-enforced trivial roots. The single-root
response is clearly an exact $\theta=\pi$ bound state, rather than a restored
continuation of the clean nontrivial root. For the double-root
target, no two distinct roots satisfy a common detuning.
For the triple-root target, the fixed root $\theta=2\pi$ remains exact
with $|\beta_2(2\pi)|=6.84\times10^{-15}$, but the required two nontrivial
additional nontrivial bound-state roots are absent under phase disorder. The survival of one isolated trivial
root therefore does not establish an exact triple-root bound
state.

\subsection{Finite-linewidth continuation of nontrivial targets}

We next consider the continuation of the clean nontrivial targets
rather than replacing them with the symmetry-enforced trivial roots.
For the same phase realization, the complex radiation-cancellation
condition is not satisfied exactly at the continued nontrivial
frequencies. We therefore compare the 

\begin{figure*}[!t]
\centering
\includegraphics[width=2.085\columnwidth]{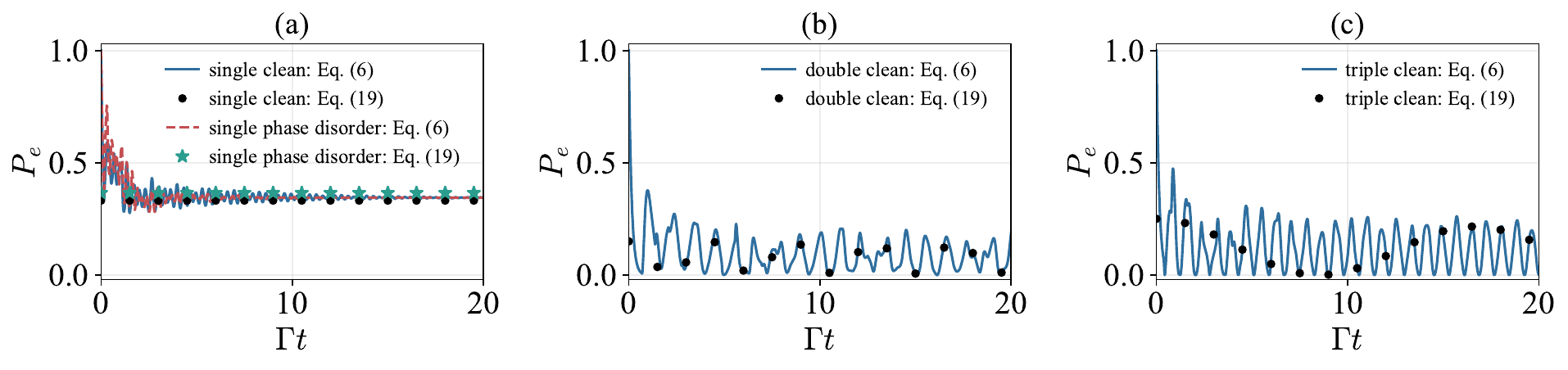}
\caption{Exact-bound-state test including the symmetry-protected trivial roots
for the seed-7 phase-disorder realization with $N=7$ and
$\eta_\phi=0.1$.
The blue solid curves and black dots represent the clean-system
excitation probability $P_e(t)=|\varepsilon(t)|^2$ from Eq.~\eqref{eq:delay-equation-general} and
bound-pole reconstructions from Eq.~\eqref{eq:common-bound-response}, respectively.
When an exact phase-disordered bound state exists, its dynamics and
reconstruction are shown by the red dashed curve and teal stars,
respectively.
Panel (a) compares the clean single-root reference with the exact
phase-disordered trivial root at $\theta_1=\pi$. Their late-time
agreement confirms the associated nondecaying contribution.
In panel (b), only the clean double-root reference is shown because no
pair of distinct phase-disordered roots satisfies a common
dispersive-matching condition.
Panel (c) shows the clean triple-root reference.
Although the symmetry-protected root $\theta=2\pi$ remains exactly
dark under phase disorder, no two nontrivial companion roots satisfy
the same matching condition. Thus, no exact phase-disordered
triple-root state is obtained.}
\label{fig:phase-trivial-root-test}
\end{figure*}

\begin{figure*}[!t]
\centering
\includegraphics[width=2.085\columnwidth]{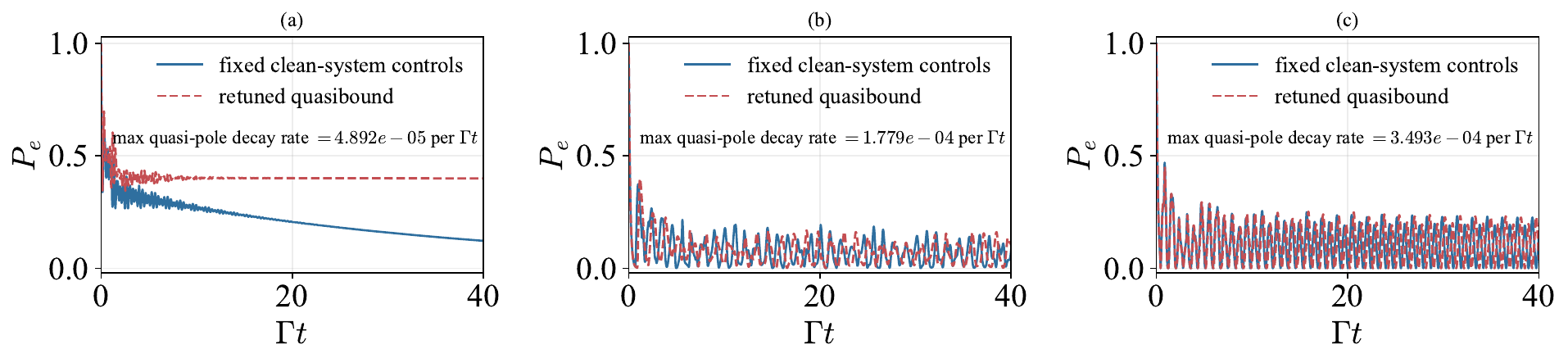}
\caption{For the seed-7 phase-disorder realization with $N=7$ and
$\eta_\phi=0.1$, (a)--(c) show the atomic excitation probability
$P_e(t)=|\varepsilon(t)|^2$ obtained from Eq.~\eqref{eq:delay-equation-general}
for the single-target, double-target, and triple-reference settings.
Blue solid curves use fixed clean-system controls, whereas red dashed
curves show retuned quasibound responses obtained by minimizing the decay
rates of the nontrivial quasipoles. In (a), $\gamma=0.0500\pi$ is
fixed and $\Delta$ changes from $1.3328\pi$ to $1.1834\pi$. In (b),
$(\gamma,\Delta)$ changes from $(0.2210\pi,1.6161\pi)$ to
$(0.2870\pi,1.3659\pi)$. In (c), the exact trivial root
$\theta=2\pi$ is retained, $\Delta=2\pi$ is fixed, and $\gamma$ changes
from $0.1250\pi$ to $0.1267\pi$ for both companion quasipoles. The maximum
probability-decay rates are $4.892\times10^{-5}$,
$1.779\times10^{-4}$, and $3.493\times10^{-4}$ per unit $\Gamma t$,
from left to right. Since the nontrivial rates remain finite, the retuned
responses are quasibound rather than exact nontrivial bound states.}
\label{fig:phase-quasibound-comparison}
\end{figure*}

After removing the common phase by setting $\phi_1=0$, the gauge-fixed
relative phases are $\phi_1=0$, $\phi_2=0.0544$,
$\phi_3=0.0301$, $\phi_4=-0.0799$, $\phi_5=-0.065$,
$\phi_6=0.0497$, and $\phi_7=-0.124$. The last relative phase
can exceed $\eta_\phi$ in magnitude because the independently sampled
raw phases lie in $[-\eta_\phi,\eta_\phi]$ before the common reference
phase is subtracted. When trivial roots are included in the root search,
the phase-disordered single-root solution is obtained at
$\theta_1\simeq\pi$, with $\gamma=0.05\pi$ and $\Delta\simeq\pi$.
The numerical residuals are $|B_\phi(\theta_1)|\simeq6.49\times10^{-11}$
and $|D_\phi(-i\theta_1)|\simeq4.36\times10^{-17}$, while
$|D_\phi'(-i\theta_1)|\simeq1.6531$. This solution therefore contains an exact bound state associated with a simple imaginary-axis pole at the trivial root $\theta_1=\pi$. Figure~\ref{fig:phase-trivial-root-test} demonstrates that phase disorder does not eliminate
the symmetry-enforced trivial roots. The single-root
response is clearly an exact $\theta=\pi$ bound state, rather than a restored
continuation of the clean nontrivial root. For the double-root
target, no two distinct roots satisfy a common detuning.
For the triple-root target, the fixed root $\theta=2\pi$ remains exact
with $|\beta_2(2\pi)|=6.84\times10^{-15}$, but the required two nontrivial
additional nontrivial bound-state roots are absent under phase disorder. The survival of one isolated trivial
root therefore does not establish an exact triple-root bound
state.
phase-disordered dynamics at
the clean values of $\gamma$ and $\Delta$ with optimized responses
in which the available parameters are chosen to minimize the decay
rates of the relevant characteristic poles. Figure~\ref{fig:phase-quasibound-comparison} shows that optimizing
$\gamma$ and $\Delta$ move

\begin{figure}[!t]
\centering
\includegraphics[width=1.01\columnwidth, height=6.9cm]{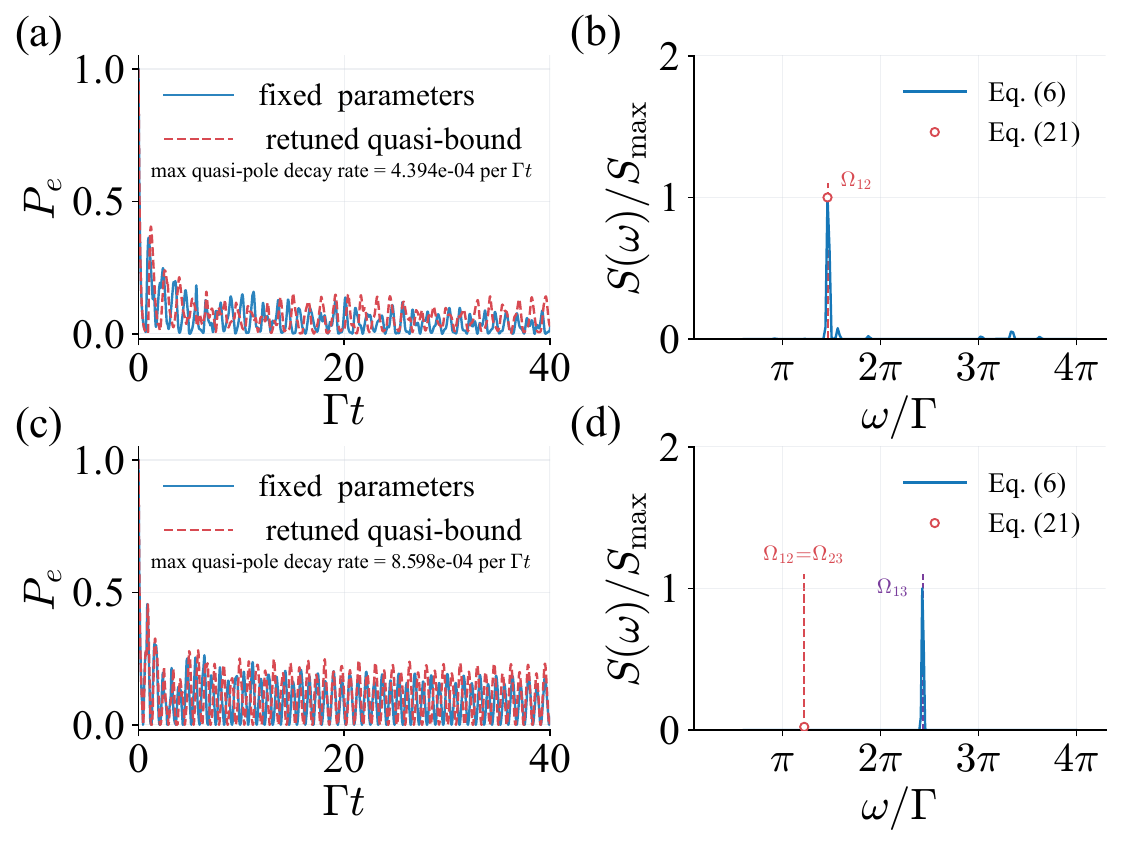}
\caption{Finite-linewidth spectral diagnostics for the phase-disordered
double-target and triple-reference responses with $N=7$,
$\eta_{\phi}=0.2$, and the seed-7 phase profile. (a) and (c)
compare the phase-disordered dynamics at the fixed clean-system values
of $\gamma$ and $\Delta$, shown by the blue solid curves, with the
retuned quasibound dynamics shown by the red dashed curves. All
time-domain curves are obtained from Eq.~\eqref{eq:delay-equation-general}. (b) and (d) show
the corresponding detrended, Hann-windowed spectra of the retuned
responses as functions of $\omega/\Gamma$, normalized by their maximum
values. The vertical dashed lines indicate the pairwise pole-beat
predictions $\Omega_{jl}^{(\mathrm{pole})}/\Gamma$ from Eq.~\eqref{eq:pole_angular_frequency},
whereas the open circles mark the nearest resolved FFT peaks evaluated
according to Eq.~\eqref{eq:spectral-diagnostics}. The retuned double-target response contains two
finite-linewidth quasipoles. The triple-reference response retains the
exact trivial root $\theta=2\pi$, while its two nontrivial companion
poles have finite linewidths.}
\label{fig:phase-beat-diagnostics}
\end{figure}

\noindent the continued nontrivial poles closer
to the imaginary axis and extends the apparent nondecaying or
oscillatory time window. It cannot, however, compensate for a
nonzero complex radiation amplitude $B_\phi(\theta)$ at the
nontrivial target frequencies. The finite negative real parts of
the optimized poles ultimately produce decay. These responses are therefore classified as quasibound continuations rather than
exactly restored nontrivial bound states. Figure~\ref{fig:phase-beat-diagnostics} (d) requires a more specific interpretation. Its
$\theta=2\pi$ component is an exact trivial bound pole, whereas the
two nontrivial companion poles retain finite linewidths. The
resulting dynamics is consequently a superposition of one exact
bound component and two decaying quasibound components, not an
exact triple-root bound state.
\subsection{Spectral and real-space signatures of bound and quasibound components}

The finite linewidths and beat frequencies are resolved using the
same seed-7 phase profile at $\eta_\phi=0.2$. For the optimized double-target response, the two nontrivial-pole
linewidths are $\vec{\kappa}^{\,(\phi)}
=(9.73\times10^{-5},\,2.04\times10^{-4})$.
The pole prediction
$f_{12}^{(\mathrm{pole})}=0.7325$ agrees with
$f^{\mathrm{FFT}}=0.729$, corresponding to
$E_{f,12}=4.7\times10^{-3}$. For the triple reference, the trivial root $\theta=2\pi$ has zero
linewidth, whereas the two nontrivial companion quasipoles have
$\kappa_1=\kappa_3=1.78\times10^{-4}$. Their degenerate predictions
$\Omega_{12}^{(\mathrm{pole})}/\Gamma
=\Omega_{23}^{(\mathrm{pole})}/\Gamma
=3.8184$
agree with the resolved FFT value
$\Omega_{\mathrm{FFT}}/\Gamma=3.834$,
corresponding to an angular-frequency error of 0.41\%. The characteristic-equation residuals remain below
$1.0\times10^{-11}$ and $1.4\times10^{-12}$ for the double and
triple cases, respectively. The agreement between the pole-predicted and resolved frequencies shows that the oscillations are governed by the continued quasipoles. A resolved beat frequency does not establish an exact bound state: the nonzero linewidths demonstrate that the
nontrivial phase-disordered components have finite lifetimes. The associated real-space fields provide the same

\begin{figure*}[!t]
\centering
\includegraphics[width=2.085\columnwidth]{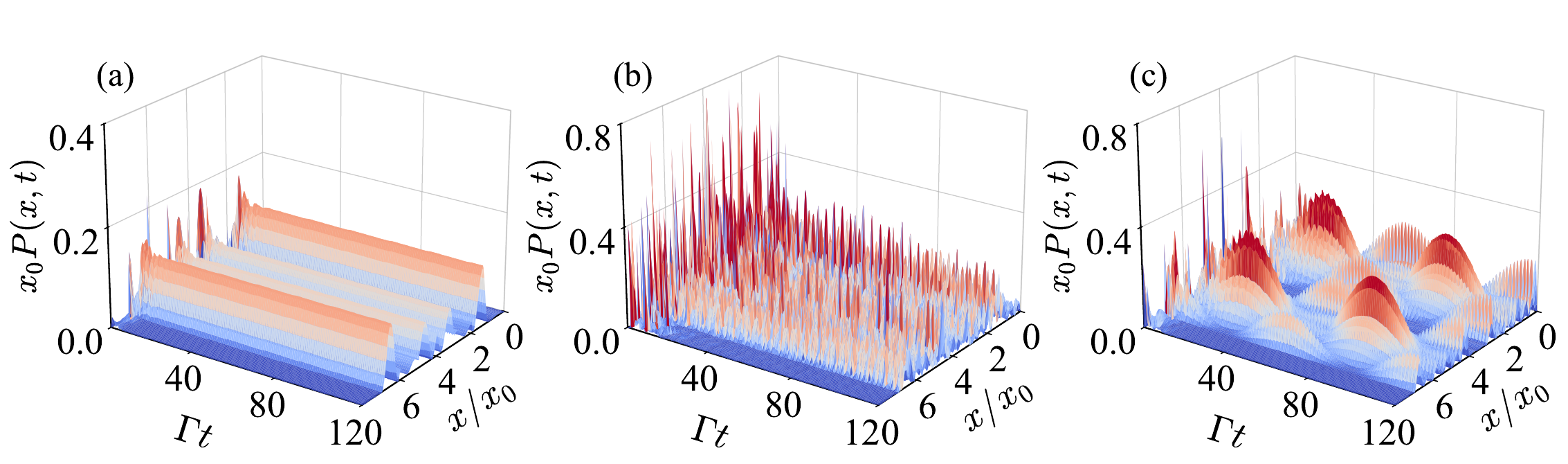}
\caption{Dimensionless real-space photon density $x_0P(x,t)$, where $P(x,t)=|\varphi(x,t)|^2$, for the optimized seed-7 phase-disorder realization with $N=7$ and $\eta_\phi=0.1$. (a)--(c) show the single-target, double-target, and triple-reference responses, respectively, calculated using the same minimum-decay parameter sets as in Figure~\ref{fig:phase-quasibound-comparison}. The corresponding maximum probability-decay rates are $4.892\times10^{-5}$, $1.779\times10^{-4}$, and $3.493\times10^{-4}$ per unit $\Gamma t$. (a) and (b) therefore show finite-lifetime quasibound photon fields. (c) contains the field generated by one exact trivial bound pole at $\theta=2\pi$ together with two finite-linewidth quasibound companion components. After the companion components decay, only the exact trivial-root contribution remains.}
\label{fig:phase-quasibound-field}
\end{figure*}

\noindent distinction. Figure~\ref{fig:phase-quasibound-field} shows finite-lifetime
single- and do\-uble-target quasibound fields. In the triple-reference
case, the two nontrivial components decay while the exact trivial
$\theta=2\pi$ contribution survives at late times. Independent coupling-phase variations rotate the radiation
contributions in the complex plane. They leave the
symmetry-enforced roots $\theta=q\pi$ exactly dark, but generically
obstruct nontrivial roots because both radiation quadratures are required to vanish simultaneously. Retuning only $\gamma$ and $\Delta$ can
recover the dispersive matching of a trivial root or reduce the
linewidths of continued nontrivial poles, but it cannot generally
restore the missing nontrivial radiation cancellation. The exact
$\theta=\pi$ and $\theta=2\pi$ components should therefore be
distinguished from the finite-linewidth nontrivial quasibound
components throughout the population, spectral, and real-space
diagnostics.

\section{Ensemble recovery statistics and mechanism comparison}
\label{sec:ensemble-statistics}

The representative realizations establish the dynamical signatures
of successful and failed matching, but they do not by themselves show
how frequently the target root sets can be recovered. We therefore
generate 1000 independent realizations for each disorder channel,
target root number, and disorder amplitude
$\eta\in\{0.05,0.10,0.15,0.20\}$. For each realization, the available
control parameters are retuned. A realization is counted as
successful only when the requested number of simple roots satisfied
the radiation-cancellation, common-detuning, and pole-residual
criteria specified in Appendix~\ref{app:numerical-protocol}. The
entries in Table~\ref{tab:ensemble-success} are ordered as single/double/triple and refer to success rates obtained under realization-dependent parameter compensation. The zero phase-channel entries refer
\FloatBarrier

\begin{table}[t]
\caption{Exact target-state recovery rates in percent for 1000
independent realizations per entry. Each triplet gives the
single/double/triple recovery rate. Trivial $n\pi$ roots are excluded
from the nontrivial phase targets.}
\label{tab:ensemble-success}
\centering
\small
\begin{ruledtabular}
\begin{tabular}{lcccc}
Channel
& $0.05$
& $0.10$
& $0.15$
& $0.20$ \\
Strength
& 100/100/100
& 100/100/100
& 100/100/99.9
& 100/99.8/99.3 \\
Position
& 100/100/0
& 100/100/0
& 100/100/0
& 100/100/0 \\
Phase
& 0/0/0
& 0/0/0
& 0/0/0
& 0/0/0 \\
\end{tabular}
\end{ruledtabular}
\end{table}
\noindent exclusively to the requested
nontrivial target sets. Symmetry-enforced roots
$\theta=q\pi$ are excluded from this statistic. Therefore, the exact
trivial single-root solution shown in
Figure~\ref{fig:phase-trivial-root-test} does not contradict the
reported $0/0/0$ nontrivial-target recovery rate. The contrast between the disorder channels follows directly from the
structure of the radiation amplitude. Strength errors preserve the
integer propagation phases and deform a real scalar cancellation
condition. Position errors also leave the cancellation equation real,
which permits the one- and two-root targets to be reconstructed over
the sampled range, but the specified triple target fails because the
fixed $\theta=2\pi$ root is no longer dark. Independent phase errors
are more restrictive because $B_\phi(\theta)=0$ contains two real
radiation-cancellation equations. Thus, no sampled phase realization
recovers the requested nontrivial single-, double-, or triple-root
target, although the symmetry-enforced trivial roots remain exact. The reported recovery rates are finite-sample numerical results obtained within the frequency windows, residual tolerances, and root-selection criteria specified in Appendix~\ref{app:numerical-protocol}. They should therefore be interpreted as evidence for the relative success of exact target-state recovery through realization-dependent compensation across the three disorder channels, rather than as a proof for all possible realizations.

\section{Conclusion}
\label{sec:conclusion}
We have derived a common criterion for atom-photon bound states of a giant atom coupled at arbitrary positions and with complex local amplitudes to a mirror-terminated waveguide. On the imaginary axis, the pole function separates into the nonnegative radiation term $\gamma|B(\theta)|^2$ and the reactive mismatch $F(\theta)$. An exact simple bound pole exists only when the same real frequency closes the outgoing channel, satisfies the atomic matching condition, and has a nonzero pole derivative. This separation also fixes the interpretation of the dynamics: one matched root produces a stationary nondecaying contribution, while two or three matched roots produce persistent pairwise beating. The three imperfection channels deform different parts of the same
interference problem. Strength errors change path weights, position
errors change both propagation phases and delays, and phase errors
rotate the radiation amplitudes in the complex plane. Representative
delay-equation solutions, pole linewidths, spectra, and real-space
fields consistently distinguish exact restoration from finite-time
trapping. The ensemble results sharpen that distinction:
realization-dependent retuning recovers the tested strength targets
almost universally and the position one- and two-root targets
universally over the sampled amplitudes, but it does not recover the
specified position triple target or any nontrivial independent-phase
target. Symmetry-enforced trivial roots $\theta=q\pi$ remain exact
when $\Delta=q\pi$, whereas the optimized responses associated with
nontrivial phase targets retain finite linewidths and are therefore
quasibound.

\FloatBarrier

Finite bandwidth, imperfect reflection, propagation attenuation, nonguided decay, and pure dephasing will generally move an ideal pole away from the imaginary axis. Quantifying those shifts requires an explicit structured-reservoir or lossy-platform calculation and is not inferred from the present delay model. At a broader platform level, engineered system-reservoir energy exchange may connect giant-atom waveguides to remote quantum-battery charging and a generalized fluctuation relation \cite{Song0904012024,Wu0504012024,Yan2026GiantAtomBattery}. Establishing either connection in the present imperfect, mirror-terminated setting requires a separate thermodynamic analysis. Within its stated scope, the radiation-dispersion decomposition provides a practical criterion for exact bound-state recovery through realization-dependent compensation and prevents long-lived transients from being misidentified as disorder-protected bound states. Looking forward, extending the present single-atom radiation-dispersion decomposition to multi-emitter configurations could reveal novel collective non-Markovian dynamics and robust chiral dark-state engineering.

\section*{AUTHOR CONTRIBUTIONS}
This project was conceived by T. Y. Liu,  H. Z. Shen. The methodology was developed by T. Y. Liu, W. Y. Hu, X. Y. Li, T. Z. Luan, Cheng Shang and H. Z. Shen. The project was supervised by C. Shang, and H. Z. Shen. The original draft of the manuscript was written by T. Y. Liu, W. Y. Hu, H. Z. Shen, and C. Shang. The manuscript was reviewed and edited by T. Y. Liu, W. Y. Hu, X. Y. Li, T. Z. Luan, Cheng Shang and H. Z. Shen.

\section*{ACKNOWLEDGMENTS}

This work was supported by Science and Technology Development Plan Project of Jilin Province (Grant No.~20250102007JC), National Natural Science Foundation of China (NSFC) under Grants No.~12274064, the Research Start-up Funds of
Shenyang Normal University (Grants No. BS202516), and Hakubi Projects of RIKEN.

\section*{CONFLICTS OF INTEREST}
The authors declare no conflicts of interest.

\section*{DATA AVAILABILITY STATEMENT}
The data that support the findings of this study are available from the corresponding author upon reasonable request.

\section*{CODE AVAILABILITY STATEMENT}
Code used to generate data in this study is available from the corresponding author upon reasonable request.\\

\noindent \textbf{Correspondence} and requests for materials should be addressed to H. Z. Shen.

\appendix 
\section{Derivation of the delay equation}
\label{app:delay-equation}

We derive the generalized delay equation by eliminating the
photonic amplitudes from the single-excitation equations of motion.
The atom is initially excited and the waveguide is taken to be in
the vacuum state, so that
$\varepsilon(0)=1$, $\varepsilon(t<0)=0$, and
$\varphi(k,0)=0$. Starting from Eqs.~\eqref{eq:eom-eps} and~\eqref{eq:eom-phi}, the
formal solution for the photonic amplitude is

\begin{equation}
\varphi(k,t)
=
-i\sum_{n=1}^{N}
g_k G_n e^{i\phi_n}\sin(kx_n)
\int_{0}^{t}dt'\,
e^{-i\omega_k(t-t')}\varepsilon(t').
\label{eq:app-photon-solution}
\end{equation}

\noindent Substitution of Eq.~\eqref{eq:app-photon-solution} into the equation for the atomic amplitude yields the memory equation $\dot{\varepsilon}(t) = -i\omega_0\varepsilon(t) - \int_{0}^{t}dt'\, K(t-t')\varepsilon(t')$. Here, the memory kernel is decomposed into pairwise contributions, $K(\sigma)=\sum_{m,n=1}^{N}K_{mn}(\sigma) (\text{with }  \sigma=t-t')$, where $K_{mn}(\sigma) = \frac{\Gamma v}{\pi} G_mG_n e^{i(\phi_n-\phi_m)} \times \int_{0}^{\infty}dk\,\sin(kx_m)\sin(kx_n) e^{-i\omega_k\sigma}$, where the linear dispersion relation
$\omega_k=vk$ is used. The factors
$G_mG_n$ and $e^{i(\phi_n-\phi_m)}$ explicitly retain the
coupling-strength and phase disorder, respectively. The product of standing-wave factors is decomposed into a direct propagation contribution and a mirror-image contribution via $\sin(kx_m)\sin(kx_n) = \frac{1}{2}\cos\bigl[k(x_m-x_n)\bigr] - \frac{1}{2}\cos\bigl[k(x_m+x_n)\bigr]$. The first cosine term corresponds to direct propagation between two
coupling points, whereas the second cosine term describes propagation
from one coupling point to the mirror and back to another coupling
point. Within the wide-band approximation, the positive-frequency
integral is extended to the full real axis. The principal-value
contribution is absorbed into the renormalized atomic frequency
$\omega_0$. The remaining resonant contribution is determined by $\int_{-\infty}^{\infty}dk\,\cos(kL)e^{-ivk\sigma} = (\pi/v)[\delta(\sigma-L/v) + \delta(\sigma+L/v)]$.
Because the memory equation is causal, only $\sigma\geq 0$ contributes.
Introducing the dimensionless coupling-point positions through
$x_m=u_m v\tau_0$, the retarded part of the kernel becomes
\begin{equation}
\begin{aligned}
K_{mn}^{(\mathrm{ret})}(\sigma)
&= \frac{\Gamma}{2} G_m G_n e^{i(\phi_n-\phi_m)} [\delta(\sigma-|u_m-u_n|\tau_0) \\
&\hspace{0.5cm} - \delta(\sigma-(u_m+u_n)\tau_0)].
\end{aligned}
\label{eq:app-retarded-kernel}
\end{equation}
The first delta function gives the direct propagation delay
$|u_m-u_n|\tau_0$, while the second gives the mirror-reflection delay
$(u_m+u_n)\tau_0$. Substituting this retarded kernel into the memory
equation yields Eq.~\eqref{eq:delay-equation-general}.

\section{Derivation of the generalized one-, two-, and
three-root bound-state conditions}
\label{app:general-bound-state-conditions}

The radiation-dispersion decomposition and the exact
bound-root conditions have already been obtained in
Eqs.~\eqref{eq:radiation-dispersion} and \eqref{eq:general-bound-condition}. We therefore derive only the parameter-compatibi\-lity conditions associated with one, two, and three distinct roots. For one root $\theta_1$, the matching relation in
Eq.~\eqref{eq:general-bound-root-set} directly fixes the
dimensionless atomic frequency $\Delta$ for a chosen positive
$\gamma$. No additional root-compatibility condition is
required. For two distinct roots $\theta_1$ and $\theta_2$, subtracting
their two matching relations gives
\begin{equation}
\gamma=
\frac{\theta_2-\theta_1}
{S(\theta_2)-S(\theta_1)}>0,
\quad
S(\theta_2)\neq S(\theta_1).
\label{eq:app-two-root-compatibility}
\end{equation}
The common value of $\Delta$ is then recovered from
Eq.~\eqref{eq:general-bound-root-set} using either root. For three distinct roots, the same subtraction applied to
the pairs $(\theta_1,\theta_2)$ and
$(\theta_1,\theta_3)$ gives
\begin{equation}
\frac{\theta_2-\theta_1}
{S(\theta_2)-S(\theta_1)}
=
\frac{\theta_3-\theta_1}
{S(\theta_3)-S(\theta_1)}
=\gamma>0,
\label{eq:app-three-root-compatibility}
\end{equation}
or
\begin{equation}
(\theta_2-\theta_1)[S(\theta_3)-S(\theta_1)]
-
(\theta_3-\theta_1)[S(\theta_2)-S(\theta_1)]
=0.
\label{eq:app-three-root-determinant}
\end{equation}
The common value of $\Delta$ is again obtained from
Eq.~\eqref{eq:general-bound-root-set}. In every case, the
retained roots are also required to satisfy the simple-pole and
normalizability requirements specified in the main text.
In the clean limit, $G_m=1$, $u_m=m$, and $\phi_m=0$, the radiation amplitude reduces to $B_0(\theta) = \sum_{m=1}^{N}\sin(m\theta) = \sin(N\theta/2)\sin[(N+1)\theta/2] / \sin(\theta/2)$. Its nontrivial dark roots belong to
\begin{equation}
\theta_k^{(N)}=\frac{2k\pi}{N},
\quad
\theta_k^{(N+1)}=\frac{2k\pi}{N+1}.
\label{eq:appendix-clean-root-families}
\end{equation}
The clean dispersive function is
\begin{equation}
S_0(\theta)
=
\sum_{d=1}^{N-1}(N-d)\sin(d\theta),
\label{eq:appendix-clean-dispersive}
\end{equation}
with
\begin{align}
S_0(\theta_k^{(N)})
&=
\frac{N}{2}
\cot(\frac{k\pi}{N}),
\label{eq:S0-N}
\\
S_0(\theta_k^{(N+1)})
&=
\frac{N+1}{2}
\cot(\frac{k\pi}{N+1}).
\label{eq:S0-N1}
\end{align}
For a selected one-, two-, or three-root configuration, we take the
roots from Eq.~\eqref{eq:appendix-clean-root-families}, evaluate the clean dispersive function using
Eqs.~\eqref{eq:S0-N} and \eqref{eq:S0-N1}, and require the selected roots to satisfy the generalized bound-state condition given by Eq.~\eqref{eq:general-bound-root-set}. The resulting equations
determine a common positive coupling parameter $\gamma$ and a common
detuning $\Delta$ for the selected roots. These values give the clean
reference parameter sets reported in the main text.

\section{Derivation and normalization of the real-space bound field}
\label{app:field-expression}

The real-space photon field follows from the formal
momentum-space solution in Eq.~\eqref{eq:app-photon-solution}.
Fourier transforming the eliminated photonic amplitude in the
standing-wave basis and separating the right- and left-moving
image contributions yields the time-domain field in
Eq.~\eqref{eq:real-space-field}. The two terms in that equation
describe direct propagation from a coupling point and propagation
through its mirror image, respectively. For a set of $M$ distinct simple imaginary-axis poles
$\{\theta_1,\ldots,\theta_M\}$, the nondecaying part of the
real-space photon field can be written as

\begin{equation}
\varphi_{\mathrm{bs}}(x,T)
=
\sum_{j=1}^{M}
\frac{\e^{-\imag\theta_jT}}
{\mathcal{N}(\theta_j)}
F_j(x),
\label{eq:appendix-field-bound}
\end{equation}
where $T=t/\tau_0$ and $\mathcal{N}(\theta_j)$ is the residue
denominator defined in
Eq.~\eqref{eq:residue-denominator-general}. The spatial mode
associated with the root $\theta_j$ is

\begin{equation}
\begin{aligned}
F_j(x)
=&
-\imag\sqrt{\frac{\Gamma}{2v}}
\sum_{m=1}^{N}
G_m\e^{\imag\phi_m}
\\
&\quad\times
[
\e^{\imag\theta_j|x-x_m|/(v\tau_0)}
-
\e^{\imag\theta_j(x+x_m)/(v\tau_0)}
],
\end{aligned}
\label{eq:appendix-field-mode}
\end{equation}
where factors $G_m$ and $\e^{\imag\phi_m}$ retain the
coupling-strength and phase imperfections explicitly, whereas
the positions $x_m=u_m v\tau_0$ retain the position
imperfections. Equation~\eqref{eq:appendix-field-bound} therefore
uses the same imperfect parameters, bound-state roots, and pole
residues as the population response in
Eq.~\eqref{eq:common-bound-response}.

\subsection{Normalization of a simple bound state}

To connect an imaginary-axis pole with a normalizable
atom-photon eigenstate, we consider a stationary
single-excitation state with energy
$E_j=\theta_j/\tau_0$ and ansatz $|\Psi_j\rangle = c_{e,j}|e,0\rangle + \int_{0}^{\infty}dk\, c_{k,j}\hat{a}_k^\dagger|\mathrm{g},0\rangle$. Projecting the stationary Schr\"odinger equat\-ion onto the atomic
and one-photon sectors gives $(E_j-\omega_0)c_{e,j} = \int_{0}^{\infty}dk\, \bigl(\mathrm{g}_{k}^{\mathrm{eff}}\bigr)^* c_{k,j}$, and $(E_j-\omega_k)c_{k,j} = \mathrm{g}_{k}^{\mathrm{eff}}c_{e,j}$, where the effective coupling between the atom and mode $k$ is $\mathrm{g}_{k}^{\mathrm{eff}} = \sum_{m=1}^{N}\mathrm{g}_{km} = \sqrt{\Gamma v / \pi} \sum_{m=1}^{N} G_m e^{i\phi_m}\sin(kx_m)$. Solving the second relation yields the photonic coefficient $c_{k,j} = \mathrm{g}_{k}^{\mathrm{eff}} c_{e,j} / (E_j-\omega_k)$. The normalization condition for the stationary eigenstate is $1 = |c_{e,j}|^2 + \int_{0}^{\infty}dk\,|c_{k,j}|^2 = |c_{e,j}|^2 [1 + \int_{0}^{\infty}dk\, |\mathrm{g}_{k}^{\mathrm{eff}}|^2 / (E_j-\omega_k)^2]$. For a simple normalizable bound state, the factor in square
brackets is, after expressing the energies in the dimensionless
convention used in the main text, equal to the derivative of the
pole denominator,
\begin{equation}
\mathcal{N}(\theta_j) = D'(z)\bigl|_{z=-\imag\theta_j} = 1+ \int_{0}^{\infty}dk\, \frac{|g_{k}^{\mathrm{eff}}|^2}{(E_j-\omega_k)^2} >0.
\label{eq:normalization-residue-relation}
\end{equation}
Consequently, the atomic weight of the normalized bound
eigenstate is

\begin{equation}
|c_{e,j}|^2
=
\frac{1}{\mathcal{N}(\theta_j)}.
\label{eq:bound-atomic-weight}
\end{equation}
This relation also explains the residue factor
$Z_j=1/\mathcal{N}(\theta_j)$ in
Eq.~\eqref{eq:common-bound-response}. For an initially excited
atom, the contribution of a normalized bound eigenstate to the
atomic survival amplitude contains both the initial overlap with
that eigenstate and its atomic component, giving the residue
weight $1/\mathcal{N}(\theta_j)$.

\subsection{Radiation cancellation and spatial confinement}

At the resonant wave number $k_j=E_j/v$, the effective coupling
to the propagating mode  is directly related to the generalized radiation amplitude via $\mathrm{g}_{k_j}^{\mathrm{eff}} = \sqrt{\Gamma v / \pi}\, B(\theta_j)$. Therefore, the radiation-cancellation condition
$B(\theta_j)=0$ removes the apparent on-shell singularity in
the expression for the photonic coefficient. The additional
dispersive condition $F(\theta_j)=0$ ensures that
$E_j=\theta_j/\tau_0$ is compatible with the atomic frequency.
Together with the simple-root condition
$\mathcal{N}(\theta_j)\neq0$, these conditions produce a
normalizable atom-photon bound state. The radiation-cancellation condition also determines the
spatial support of the bound photon field. For $x>x_N$, where $x_N$ is the outermost coupling point, one has
$|x-x_m|=x-x_m$ for every $m$, so Eq.~\eqref{eq:appendix-field-mode} reduces to $F_j(x) = -i\sqrt{\Gamma/(2v)}\, e^{i\theta_j x/(v\tau_0)} \sum_{m=1}^{N} G_m e^{i\phi_m} [e^{-i u_m\theta_j} - e^{i u_m\theta_j}] = -2\sqrt{\Gamma/(2v)}\, e^{i\theta_j x/(v\tau_0)} B(\theta_j)$.
It follows immediately that
\begin{equation}
B(\theta_j)=0
\quad\Longrightarrow\quad
F_j(x)=0
\quad
(x>x_N).
\label{eq:bound-field-compact-support}
\end{equation}
Thus, an exact bound-state photon field is confined to the finite
region between the mirror and the outermost coupling point. The
radiation condition $B(\theta_j)=0$ prevents emission into the
open part of the waveguide, while the dispersive condition
$F(\theta_j)=0$ places the pole at the same real frequency as the
atom-photon eigenstate. Equations~\eqref{eq:appendix-field-bound} and
\eqref{eq:appendix-field-mode} establish the relation between the
real-space field plots and the pole residues. A single
imag\-inary-axis pole produces a stationary trapped density
profile. Two distinct imaginary-axis poles produce a
time-dependent interference pattern with dimensionless beat
frequency $|\theta_1-\theta_2|$, and three distinct poles produce
up to three pairwise beat frequencies
$|\theta_j-\theta_\ell|$. The exact decomposition in Eq.~\eqref{eq:appendix-field-bound} contains only imagi\-nary-axis poles satisfying Eq.~\eqref{eq:general-bound-condition}. A continued pole with a
finite negative real part instead produces a decaying
real-space component and represents a quasibound transient,
rather than a normalizable stationary bound eigenstate. This
distinction is used when interpreting the optimized
phase-imperfection photon fields.

\section{Reality of the bound-state frequency parameter}
\label{app:theta-classification}

Following the imaginary-axis parametrization introduced immediately
after Eq.~\eqref{eq:pole-general}, we write
\(z=-\imag\theta\). It is important to distinguish the position of
the pole \(z\) from the character of the parameter \(\theta\). In
particular, the statement that a bound-state pole lies on the
imaginary axis does not mean that \(\theta\) is imaginary. On the
contrary, an imaginary-axis pole is obtained when \(\theta\) is real. To make this distinction explicit, write a general complex value as
\(\theta=\theta_R+i\theta_I\), with \(\theta_R,\theta_I\in\mathbb{R}\).
Using this parametrization in \(z=-i\theta\) gives
\(z=\theta_I-i\theta_R\), so that
\(\operatorname{Re}z=\theta_I\) and
\(\operatorname{Im}z=-\theta_R\).
The corresponding pole contribution to the time-domain excitation
amplitude follows from the residue expansion in Eq.~\eqref{eq:residue-sum} as $e^{zT} = e^{\theta_{\mathrm{I}} T} e^{-i\theta_{\mathrm{R}} T}$ (with \(T=t/\tau_0\)). Its magnitude is
\begin{equation}
\left|\e^{zT}\right|
=
\e^{\theta_{\mathrm I}T}.
\label{eq:complex-theta-magnitude}
\end{equation}
The relation \(\operatorname{Re}z=\theta_{\mathrm I}\), together with
Eq.~\eqref{eq:complex-theta-magnitude}, distinguishes the following
three cases.

\subsection{Real \texorpdfstring{$\theta$}{theta}}

If \(\theta=\theta_{\mathrm R}\in\mathbb R\), then
\(\theta_{\mathrm I}=0\), and hence
\(z=-\imag\theta_{\mathrm R}\) with \(\operatorname{Re}z=0\). The pole
therefore lies on the imaginary axis. The exponential factor is
therefore \(\e^{-\imag\theta_{\mathrm R}T}\), identical to the factor
carried by each individual bound-pole term in
Eq.~\eqref{eq:common-bound-response}, and it has unit modulus. Thus, a real value of \(\theta\) does not produce exponential decay.
It is the only case that can represent the frequency of an exact
stationary atom--photon bound state. Equivalently, the corresponding
physical energy is $E = \theta_{\mathrm{R}} / \tau_0$, which is purely real as required for a normalizable eigenstate of the full Hermitian atom-waveguide Hamiltonian. However, the reality of \(\theta\) alone is not sufficient. The
candidate value needs to also satisfy \(D(-\imag\theta)=0\), where \(D(z)\)
is the dimensionless characteristic function defined in
Eq.~\eqref{eq:pole-general}. For real \(\theta\), the
radiation--dispersion decomposition in
Eq.~\eqref{eq:radiation-dispersion} applies. Since \(\gamma>0\), its
vanishing is equivalent to the two conditions in
Eq.~\eqref{eq:general-bound-condition}. The first condition cancels
the outgoing radiation channel, whereas the second matches the
dispersive shift to the atomic frequency. The retained pole is additionally simple.
Equatio\-n\-~\eqref{eq:residue-denominator-general} identifies
\(\mathcal N(\theta)=D'(z)|_{z=-\imag\theta}\), and the admissible-root
set in Eq.~\eqref{eq:general-bound-root-set} requires
\(\mathcal N(\theta)\neq0\). The associated atom--photon eigenstate also has a finite normalization. Consequently, a real
\(\theta\) is a necessary but not sufficient condition for an exact
bound state.

\subsection{Purely imaginary \texorpdfstring{$\theta$}{theta}}

For a nonzero purely imaginary value, write
\(\theta=\imag\chi\), where
\(\chi\in\mathbb R\setminus\{0\}\). The relation
\(z=-\imag\theta\) then gives \(z=\chi\in\mathbb R\), and the
time-dependent factor becomes $e^{zT} = e^{\chi T}$. If \(\chi<0\), the pole lies on the negative real axis and gives a
nonoscillatory exponentially decaying contribution. Such a pole
describes a transient relaxation mode and not an exact bound state.
If \(\chi>0\), the contribution grows exponentially. This would
indicate an instability and is excluded for the passive system
considered here. The special value \(\chi=0\) gives \(\theta=0\) and \(z=0\).
Although zero can formally be regarded as both real and purely
imaginary, this case belongs to the real-\(\theta\) boundary and can
represent a bound state only if it separately satisfies the
admissible-root requirements in
Eq.~\eqref{eq:general-bound-root-set} and the normalization condition.
Therefore, a nonzero purely imaginary \(\theta\) cannot describe an
exact bound state.

\subsection{General complex \texorpdfstring{$\theta$}{theta}}

For a genuinely complex value
\(\theta=\theta_{\mathrm R}+\imag\theta_{\mathrm I}\), with
\(\theta_{\mathrm R}\neq0\) and \(\theta_{\mathrm I}\neq0\), the
corresponding pole has
\(\operatorname{Re}z=\theta_{\mathrm I}\neq0\). For a stable decaying
resonance, \(\theta_{\mathrm I}<0\), and the pole contribution is $e^{zT} = e^{-|\theta_{\mathrm{I}}|T} e^{-i\theta_{\mathrm{R}}T}$. It therefore represents a finite-linewidth oscillatory resonance or
a quasibound transient. Its oscillation frequency is determined by
\(\theta_{\mathrm R}\), while its amplitude-decay rate is
\(|\theta_{\mathrm I}|\). It is not a stationary, normalizable bound
state because its pole does not lie on the imaginary axis. The radiation--dispersion decomposition in
Eq.~\eqref{eq:radiation-dispersion} was derived for real \(\theta\)
and should not be applied unchanged to a complex \(\theta\). General
complex resonances are instead obtained by solving \(D(z)=0\)
with the full characteristic function in
Eq.~\eqref{eq:pole-general}, where
\(z=\theta_{\mathrm I}-\imag\theta_{\mathrm R}\).

\subsection{Bound-state criterion}

The preceding cases can be summarized as
\begin{align}
\theta\in\mathbb R
&\quad\Longleftrightarrow\quad
\operatorname{Re}z=0, \notag \\
\operatorname{Im}\theta<0
&\quad\Longleftrightarrow\quad
\operatorname{Re}z<0, \label{eq:theta-pole-summary} \\
\operatorname{Im}\theta>0
&\quad\Longleftrightarrow\quad
\operatorname{Re}z>0. \notag
\end{align}Accordingly, only a real value of \(\theta\) can place the pole on
the imaginary axis and hence qualify as a candidate exact bound
state. The complete criterion is therefore the admissible
real-frequency root condition encoded in
Eq.~\eqref{eq:general-bound-root-set}, where
\(\Delta=\theta-\gamma S(\theta)\) is equivalent to
\(F(\theta)=0\), together with finite atom--photon normalization. A
purely imaginary or general complex nonreal value of \(\theta\)
produces a pole away from the imaginary axis and therefore describes
decay, growth, or a finite-linewidth quasibound response rather than
an exact bound state.

\section{Numerical protocol and generalized root-set construction}
\label{app:numerical-protocol}

\subsection{Time-domain integration and spectral analysis}

The generalized delay-differential equation in
Eq.~\eqref{eq:delay-equation-general} was integrated using a
fourth-order Runge--Kutta method on a uniform grid in \(\Gamma t\).
The time step was$\Delta(\Gamma t)=0.02$ and the causal prehistory was set to
\(\varepsilon(t<0)=0\). Delayed amplitudes that did not coincide
with stored grid points were evaluated by interpolation from the
causal numerical history. Unless a different value is stated in a
figure caption, the final integration time was \(\Gamma t=300\). For each realization, the same sampled arrays
\(\{G_m^{(s)}\}\), \(\{u_m^{(s)}\}\), and
\(\{\phi_m^{(s)}\}\), generated from
Eq.~\eqref{eq:disorder-realizations}, were used in the delay equation,
Eq.~\eqref{eq:delay-equation-general}, the characteristic function,
Eq.~\eqref{eq:pole-general}, the residue denominator,
Eq.~\eqref{eq:residue-denominator-general}, and the real-space field,
Eq.~\eqref{eq:real-space-field}. The time-domain, pole, residue, and
field calculations therefore describe the same imperfect physical
configuration. Before evaluating a population spectrum, the selected finite-duration time trace was linearly detrended and multiplied by a Hann window. The time interval
and frequency interval used for each spectrum are stated in the
corresponding figure caption. Consistent with
Sec.~\ref{sec:finite-linewidth-diagnostics}, a spectral peak was used
only as a dynamical diagnostic. An exact bound-state component was
identified through the physical root conditions in
Eq.~\eqref{eq:general-bound-condition}, the simple-pole requirement
in Eq.~\eqref{eq:residue-denominator-general}, and the normalization relations derived above and in
\eqref{eq:normalization-residue-relation}. The absence of an outgoing
field beyond the outermost coupling point then follows from
Eq.~\eqref{eq:bound-field-compact-support}.

\subsection{Root search and numerical acceptance}

For a fixed realization $s$, the characteristic function $D^{(s)}(z)$,
radiation amplitude $B^{(s)}(\theta)$, dispersive function $S^{(s)}(\theta)$, and matching function $F^{(s)}(\theta)$ are obtained from Eqs.~\eqref{eq:pole-general}-\eqref{eq:general-match-revised} respectively, using the same sampled parameters. Their strength-, position-, and phase-imperfection specializations are given in
Eqs.~\eqref{eq:strength-root-test}-\eqref{eq:phase-dispersive-shift}, and
these analytical expressions are not repeated here. The positive-frequency search started at
\(\theta_{\min}=10^{-7}\). Each calculation used its recorded
channel- and multiplicity-dependent endpoint
\(\theta_{\mathrm{scan}}\) and a dense uniform grid. For the real
radiation amplitudes in the strength and position calculations, the
code retained sampled points satisfying
\(\left|B^{(s)}(\theta)\right|\leq10^{-12}\) and bracketed every
sign-changing interval. Each bracket was refined by Brent's method
with absolute and relative tolerances \(10^{-13}\) and at most
\(200\) iterations. Real roots separated by less than
\(10^{-10}\max\{1,|\theta|\}\) were merged. For phase imperfections, the radiation amplitude in
Eq.~\eqref{eq:phase-root-test} is generally complex, so a
sign-changing bracketing method is not applicable. The code located
grid-resolved local minima of
\(\lvert B^{(s)}(\theta)\rvert^2\) and refined each minimum by bounded
scalar minimization with an absolute frequency tolerance of
\(10^{-13}\) and at most \(200\) iterations. A refined minimum was
retained as a root candidate only when
\(\lvert B^{(s)}(\theta)\rvert^2\leq10^{-12}\). Complex-amplitude
candidates separated by less than
\(10^{-7}\max\{1,|\theta|\}\) were merged. All retained candidates
were sorted in increasing order. For realization \(s\), \(r_B^{(s)}(\theta)\) and
\(r_F^{(s)}(\theta)\) denote the radiation and matching residuals,
respectively, at a candidate frequency:
\begin{equation}
r_B^{(s)}(\theta)
=
|B^{(s)}(\theta)|,
\quad
r_F^{(s)}(\theta)
=
|F^{(s)}(\theta)|.
\label{eq:app-root-residuals}
\end{equation}
The independently evaluated pole residual and derivative magnitude are
denoted by \(r_D^{(s)}(\theta)\) and \(r_{D'}^{(s)}(\theta)\),
respectively:
\begin{equation}
r_D^{(s)}(\theta)
=
|D^{(s)}(-\imag\theta)|,
\quad
r_{D'}^{(s)}(\theta)
=
|D^{(s)\prime}(-\imag\theta)|,
\label{eq:app-pole-acceptance}
\end{equation}
where \(D^{(s)\prime}(z)=dD^{(s)}(z)/dz\), so that
\(r_{D'}^{(s)}(\theta)\) is the derivative magnitude at the candidate
pole \(z=-\imag\theta\).
The numerical tolerances were
\begin{equation}
\epsilon_B
=
\epsilon_F
=
\epsilon_D
=
10^{-7},
\quad
\epsilon_{D'}
=
10^{-10}.
\label{eq:app-numerical-tolerances}
\end{equation}
A resolved candidate was accepted only when $r_B^{(s)}\leq\epsilon_B$, $r_F^{(s)}\leq\epsilon_F$, $r_D^{(s)}\leq\epsilon_D$, and $r_{D'}^{(s)}>\epsilon_{D'}$, the first two inequalities numerically implement
Eq.~\eqref{eq:general-bound-condition}. The derivative test is the
numerical simple-pole condition associated with
Eq.~\eqref{eq:residue-denominator-general}. The quantity
\(r_D^{(s)}\) is an independently implemented consistency check, not
a third physical condition, because
Eq.~\eqref{eq:radiation-dispersion} expresses
\(D(-\imag\theta)\) entirely in terms of \(B(\theta)\) and
\(F(\theta)\). Only roots satisfying the normalization conditions derived above were retained. For comparison with the exact root set in
Eq.~\eqref{eq:general-bound-root-set}, we denote its positive-frequency
sector by
\begin{equation}
\mathcal R_{\mathrm B}^{+,(s)}(\gamma,\Delta)
=
\mathcal R_{\mathrm B}^{(s)}(\gamma,\Delta)
\cap(0,\infty),
\label{eq:app-positive-root-sector}
\end{equation}
where the superscript \(s\) means that the functions entering
Eq.~\eqref{eq:general-bound-root-set} are evaluated for realization
\(s\). The retained realizations satisfy \(G_m^{(s)}\geq0\), as specified
after Eq.~\eqref{eq:disorder-realizations}. The total coupling strength
of realization \(s\) is
\(G_\Sigma^{(s)}=\sum_{m=1}^{N}G_m^{(s)}\). The triangle inequality
applied to Eq.~\eqref{eq:general-matching-full} gives
\begin{equation}
|S^{(s)}(\theta)|
\leq
[G_\Sigma^{(s)}]^2.
\label{eq:app-dispersive-search-bound}
\end{equation}
For \(\gamma>0\), the matching relation in
Eq.~\eqref{eq:general-match-revised} therefore implies
\[
0<\theta
\leq
|\Delta|
+
\gamma
[G_\Sigma^{(s)}]^2
\]
for every positive exact root. The validation endpoint was chosen as
\begin{equation}
\theta_{\mathrm{val}}^{(s)}(\gamma,\Delta)
=
\max
\{
\theta_{\mathrm{scan}},
|\Delta|
+
\gamma
[G_\Sigma^{(s)}]^2
+\pi
\}.
\label{eq:app-validation-window}
\end{equation}
When the validation end point exceeded the primary endpoint, the scan
was repeated on the enlarged interval while preserving at least the
grid density of the primary scan. The resulting finite-window, finite-tolerance numerical root set is
\begin{equation}
\begin{split}
    \mathcal{R}_{\mathrm{B,val,num}}^{+,(s)}(\gamma, \Delta) &= \{ \theta \in [10^{-7}, \theta_{\mathrm{val}}^{(s)}] : \\
    &\quad r_B^{(s)}(\theta) \le \epsilon_B, \;\quad r_F^{(s)}(\theta) \le \epsilon_F, \\
    &\quad r_D^{(s)}(\theta) \le \epsilon_D, \;\quad r_{D'}^{(s)}(\theta) > \epsilon_{D'} \}.
\end{split}
\label{eq:app-windowed-accepted-root-set}
\end{equation}
A member of
\(\mathcal R_{\mathrm B,\mathrm{val},\mathrm{num}}^{+,(s)}\) was
retained only when its associated mode also satisfied the
normalization conditions cited above. The real-amplitude procedure is a finite sign scan followed by Brent
refinement. It detects resolved sign-changing roots and sampled
near-zero points, but it cannot exclude an isolated zero that neither
changes sign nor coincides with a sampled point. An unresolved
even-multiplicity or tangential zero of a real radiation amplitude
may therefore be missed. The numerical cardinalities below are
strict with respect to the resolved set in
Eq.~\eqref{eq:app-windowed-accepted-root-set}, rather than a proof
that no unresolved sign-preserving zero exists.

\subsection{Construction of admissible root sets}

For every realization, the sampled arrays in Eq.~\eqref{eq:disorder-realizations} were
held fixed throughout the construction of all corresponding
control solutions. Only the global controls contained in the allowed
control subspace $\mathcal{P}_{\mathrm{g}}^{(s)}$ were varied, while the
root frequencies were outputs of the root equations. Candidate control values for one root were obtained directly from
Eq.~\eqref{eq:general-bound-root-set}. Two-root candidates were
constructed with Eq.~\eqref{eq:app-two-root-compatibility}, and
three-root candidates were constructed with
Eqs.~\eqref{eq:app-three-root-compatibility} and
\eqref{eq:app-three-root-determinant}. These analytical
compatibility relations are derived in
Appendix~\ref{app:general-bound-state-conditions} and are not
rederived here. If a particular calculation imposed an additional
symmetry constraint, that constraint was included in
\(\mathcal P_M^{(s)}\), but it was not introduced as a
frequency-ranking rule. A candidate solution is written as
\begin{multline}
Q_{\alpha,M}^{(s)}
=
\bigl(
\vec{\theta}_{\alpha,M}^{\, (s)},
\gamma_{\alpha,M}^{(s)},
\Delta_{\alpha,M}^{(s)}
\bigr),\\
0<\theta_{\alpha,M,1}^{(s)}
<\cdots<
\theta_{\alpha,M,M}^{(s)}.
\label{eq:app-candidate-solution}
\end{multline}
The complete admissible collection is

\begin{equation}
\begin{aligned}
\mathcal{C}_M^{(s)} 
&= \{ \mathcal{Q}_{\alpha,M}^{(s)} : 
(\gamma_{\alpha,M}^{(s)}, \Delta_{\alpha,M}^{(s)}) \in \mathcal{P}_M^{(s)}, \\
&\quad \mathcal{R}_{\mathrm{B},\mathrm{val},\mathrm{num}}^{+,(s)} 
(\gamma_{\alpha,M}^{(s)}, \Delta_{\alpha,M}^{(s)}) \\
&\quad = \{ \theta_{\alpha,M,1}^{(s)}, \dots, \theta_{\alpha,M,M}^{(s)} \} \}.
\end{aligned}
\label{eq:app-admissible-root-sets}
\end{equation}
Thus, membership in \(\mathcal C_M^{(s)}\) requires the selected
frequencies to be exactly the complete resolved root set for the same
controls. Numerically duplicate solutions were merged after their
ordered roots and controls had been placed in a canonical order, and
all distinct members were retained. No independently assigned frequency, distance-to-refer\-ence functional, or
root-spacing score enters Eq.~\eqref{eq:app-admissible-root-sets}.
The frequencies are determined by the bound-root equations rather
than assigned independently. The number of admissible $M$-root solutions for realization $s$, denoted by $K_M^{(s)}$, is given by
\begin{equation}
K_M^{(s)}
=
\operatorname{card}\mathcal C_M^{(s)}.
\label{eq:app-admissible-solution-count}
\end{equation}
A realization was classified as viable precisely when
\(K_M^{(s)}>0\).
If one representative was needed for a plot or a compact numerical
record, the members of \(\mathcal C_M^{(s)}\) were first sorted
lexicographically by their ordered roots and controls. One index was
then sampled uniformly,
\begin{equation}
q_M^{(s)}
\sim
\operatorname{Uniform}
\{
1,\ldots,K_M^{(s)}
\},
\quad
\mathcal Q_{\mathrm{rep},M}^{(s)}
=
\mathcal Q_{q_M^{(s)},M}^{(s)}.
\label{eq:app-uniform-representative-selection}
\end{equation}
This draw selects one already admissible solution, but it does not
assign, perturb, or randomly prescribe a root frequency. It does not
change \(\mathcal C_M^{(s)}\), \(K_M^{(s)}\), or any ensemble success
statistic.

\subsection{Complex-pole continuation and phase-linewidth optimization}

The exact-root construction and the finite-linewidth phase
optimization were treated as separate calculations. Exact
phase-disordered solutions were required to belong to
\(\mathcal C_M^{(s)}\). Finite-linewidth solutions were not inserted
into the exact admissible collection. Complex poles were obtained by solving the real and imaginary parts
of \(D^{(s)}(z)=0\), with \(D^{(s)}(z)\) defined in
Eq.~\eqref{eq:pole-general}. Each solve was initialized near a clean
pole or near the pole obtained at the preceding continuation step.
The nonlinear-solver tolerance was \(10^{-11}\), and a pole was
accepted only when
\begin{equation}
|D^{(s)}(z_j)|<10^{-7}.
\label{eq:app-complex-pole-residual}
\end{equation}
Multiple initial offsets were tested, numerical duplicates were
merged, and the closest one-to-one continuation was retained. For a
pole on the imaginary axis, the derivative was additionally evaluated
using Eq.~\eqref{eq:residue-denominator-general}. Using the continued-pole notation and time scaling introduced in
Sec.~\ref{sec:finite-linewidth-diagnostics}, the probability-decay
rate of pole \(j\) per unit \(\Gamma t\) is
\(2\kappa_j/\gamma\). When an exact nontrivial phase-disordered set
was absent, the available controls were varied to minimize
\begin{equation}
J_\phi( p)
=
\max_{j\in\mathcal T_{\mathrm{nt}}}
\frac{2\kappa_j(p)}{\gamma},
\label{eq:app-phase-linewidth-objective}
\end{equation}
where \(\mathcal T_{\mathrm{nt}}\) contains the nontrivial poles
continued from one clean admissible solution. The optimization variables were
\[
p
=
\Biggl\{
\begin{array}{@{}l@{ }l@{}}
(\Delta),
&
\hspace{-1em}\text{one-root calculation with fixed \(\gamma\),}
\\
(\gamma,\Delta),
&
\text{two-root calculation,}
\\
(\gamma),
&
\hspace{-1em}\text{three-root calculation retaining a symmetry root.}
\end{array}
\]
In the last case, \(\Delta\) was held at the matching value required
by Eq.~\eqref{eq:general-match-revised} for the retained symmetry
root, while \(\gamma\) was varied to reduce the linewidths of the
nontrivial companion poles. If several clean admissible solutions
were available, the solution used for this representative
calculation was selected once by
Eq.~\eqref{eq:app-uniform-representative-selection} before
continuation began. An optimized response was classified as quasibound whenever at least
one retained nontrivial pole had \(\kappa_j>0\). Such a response was
not counted as an exact member of \(\mathcal C_M^{(s)}\).

\subsection{Ensemble sampling and scope of the numerical classification}

The ensemble calculations used realization indices
\(s=0,\ldots,999\). The strength, position, raw phase, and
represen\-tative-selection streams were initialized using
\begin{equation}
\begin{aligned}
s_G&=s, &
s_x&=10^6+s,\\
s_\phi&=2\times10^6+s, &
s_{\mathrm{sel}}&=3\times10^6+s.
\end{aligned}
\label{eq:app-ensemble-seed-convention}
\end{equation}
The first three streams generate the disorder variables in
Eq.~\eqref{eq:disorder-realizations}. The fourth stream is independent
of the physical-disorder generation and is used only for
Eq.~\eqref{eq:app-uniform-representative-selection}. Bitwise
reproduction additionally requires the same random-number-generator
implementation and software version. For every realization, all members of \(\mathcal C_M^{(s)}\) were
retained. The success criterion used for the ensemble classification
is Eq.~\eqref{eq:app-admissible-solution-count}, which is independent of
the representative index. Finite-linewidth quasibound optimizations
were excluded from the exact success count. The classification is numerical rather than a global theorem. The
validation endpoint in Eq.~\eqref{eq:app-validation-window} bounds
the positive roots allowed by the matching condition, but the real
sign scan retains the unresolved sign-preserving-zero limitation
stated above, and the complex minimum scan cannot exclude a minimum
narrower than the sampling resolution. The reported card\-inalities
therefore refer to the resolved numerical sets defined in
Eq.~\eqref{eq:app-windowed-accepted-root-set} under the stated
tolerances. Finally, the transcendental characteristic Eq.~\eqref{eq:pole-general} has infinitely many complex roots. The
complex-pole procedure follows only the poles connected to the
selected clean poles or to poles found at preceding continuation
steps. It is not a global enumeration of all far-left complex roots,
which decay rapidly and do not contribute to the asymptotic
bound-pole reconstruction.

\end{document}